\documentclass[lettersize,journal]{IEEEtran}
\usepackage{amsmath,amsfonts}
\usepackage{amssymb}
\usepackage{algorithm}
\usepackage{algpseudocode}
\algrenewcommand\algorithmiccomment[1]{\Statex \(\#\) #1}
\usepackage{array}
\usepackage{textcomp}
\usepackage{stfloats}
\usepackage{url}
\usepackage{verbatim}
\usepackage{graphicx}
\usepackage{subcaption}
\usepackage{float}
\usepackage[]{mdframed}
\usepackage{mathtools}
\usepackage{comment}
\usepackage{cite}
\usepackage{stmaryrd}
\usepackage{makecell}
\usepackage{amsthm}
\newtheorem{theorem}{Theorem}[section]
\newtheorem{corollary}{Corollary}[theorem]
\newtheorem{lemma}[theorem]{Lemma}

\newtheorem{definition}{Definition}[section]

\makeatletter
\newcommand{\vo}{\vec{o}\@ifnextchar{^}{\,}{}}
\makeatother
\begin{document}

\title{PSA: Private Set Alignment for Secure and Collaborative Analytics on Large-Scale Data}

\author{Jiabo Wang, Elmo Xuyun Huang, Pu Duan, Huaxiong Wang, Kwok-Yan Lam

 \thanks{Dr Jiabo Wang is with SCRiPTS, Nanyang Technological University, Singapore. Dr Pu Duan is with Ant International. Prof Huaxiong Wang is with the Division of Mathematical Sciences, Nanyang Technological University, Singapore. Prof Kwok-Yan Lam and Mr Elmo Xuyun Huang are with the College of Computing and Data Science, Nanyang Technological University, Singapore.}}

\maketitle

\begin{abstract}

As an important part of data innovation, organizations are tempted to join forces in pooling their data and performing analytics so as to enable them to have better insight of market and customer behavior trends, and more effective ability to detect anomalous and even fraudulent activities in the industry ecosystem. Whereas such efforts could potentially lead to breach of data privacy policies and regulations. To allow collaborative analytics while complying with the applicable privacy regulations, Privacy-Enhancing Technologies (PET) are widely believed to be a promising area that attracted the attention of academic researchers and industry practitioners. In this work, we developed a technique that addresses scenarios where two organizations collaborate to perform data analytics by jointly using their respective data but without sharing the data with each other. The two organizations are required to securely identify common customers/users and virtually inner join their datasets with respect to these customers, and to perform analytics on this virtual dataset in order to maximize data insights. Apart from the mandatory protection of raw customer data, it becomes more challenging to protect identities and attributes of common customers, as it requires participants to align their records associated with common customers without knowing who they are. We proposed a solution, namely Private Set Alignment (PSA), for this scenario, which is proven to be effectively applicable to real-world use cases, such as evaluating conversion rate of advertisement using data from both advertisement distribution channels (publishers) and advertisers (merchants). 

The contributions of this work are threefold. Firstly, we defined the notion of PSA with two levels of privacy protection and designed the protocols to achieve them. Particularly, at the core of our protocols is a modified oblivious switching network which leverages efficient symmetric key operations and offline pre-computation to save online run time. Secondly, we implemented and benchmarked the proposed protocols in different network conditions. For comparison of experimental results, our method joins two datasets, each at the scale of one million records, in 35.5 sec on a single thread with a network bandwidth of 500 Mbps, while an existing technique, which is based homomorphic encryption, takes approximately 56.3 min in the same setting, resulting in × 100 improvement. Thirdly, we give a new proof for the existence of an algorithm of quasi-linear complexity to construct an oblivious switching network to achieve a target permutation, as is distinct from the existing one in the literature.

\end{abstract}


\section{Introduction}
\IEEEPARstart{C}{o}llaborative data analytics has shown its amazing power to help enterprises, academia and government agencies uncover the valuable insights beneath the data, leading to some of the most successful businesses, new technologies, and best ever public services. However, due to regulatory requirements and privacy
concerns, 
the data owners more than ever underscore the custody of the data they harvested and sorted out from their business and activities. This, on the contrary, builds up hurdles for collaborative data analytics across peer organizations. We have seen privacy-preserving frameworks of different levels of protection to facilitate analytics on data from distributed sources, e.g. Secrecy \cite{liagouris2023secrecy}, Conclave \cite{volgushev2019conclave}, Senate\cite{poddar2021senate}, SMCQL \cite{bater2017smcql}, Shrinkwrap \cite{bater2018shrinkwrap}, SecretFlow \cite{secretflow23}, etc.

In this work, we developed a technique, namely Private Set Alignment (PSA), to provide a solution for efficiently and securely joining distributed datasets with respect to common customers/users. In this scenario, there exist two structured datasets (tables) each describing certain attributes associated with customer/user IDs, and the two tables share the same ID system to identify their customers/users. 
PSA is a secure computation protocol between two data owners that takes the two tables as input and returns the secret shares of the records associated with the shared IDs only (i.e., inner join). It is privacy-preserving because neither participants learns the other's dataset, the inner join, or even the shared IDs in clear. Particularly, the inner join, as protocol output, is in the format of secret shares such that no participants learns it in clear unless they agree to work together to reveal it. Moreover, the secret share format provides interface to subsequent MPC analytics on the inner join of the two tables. 

 As a high level description, the PSA protocol proceeds with two steps: securely identifying the matched IDs then aligning their associated records to create a new table. Depending on the information leakage amid the first step, we in this work specifically define PSA in 2 privacy settings (a) \textit{single blinded} when one participant is allowed to learn the matched IDs (b) \textit{double blinded} when neither participants learns the matches. 



We have identified many real-world use cases that would benefit from PSA. A typical one is for advertising conversion analytics in e-commerce \cite{ion2020deploying}. The publishers (e.g., social media APPs, streaming websites) know customers browsing history all the while the merchants know customers purchasing activities. Combining and analyzing the data from the two parties can help calculate the advertising revenue and improve the advertising strategies. Because PSA returns inner join in secret shares, it provides interface to privacy-preserving machine learning (PPML) via MPC \cite{secureML17}, enabling model training with distributed data. A use case of this kind is heart failure prediction \cite{LANCELOT2020}. In this use case, a medical centre has data of people's lifestyle (e.g., smoking, exercising behaviors) in a city while a medical insurance company claims people's hospitalization data. They compute the inner join of the two datasets in a privacy-preserving way and train a regression model with the combined data for prediction by PPML. Additionally, PSA holds promise for facilitating use cases such as anti-money laundering and contact tracing during a pandemic, where securely and efficiently joining distributed datasets is essential
\cite{antiLaund24,duong2020catalic,reichert2021circuit}.

There are existing works to achieve similar functionality to PSA, though different privacy requirements may apply. Some are built on top of MPC to provide secure and versatile oblivious queries on big data from multiple sources \cite{volgushev2019conclave}. Some employ ``custom-designed'' protocols to provide efficient realization of certain types of oblivious queries, e.g. SUM of an attribute associated with the intersection \cite{ion2020deploying}. But when generalized to oblivious queries of arbitrary types, those custom protocols often lose their advantages in efficiency or even capability. A framework that leverages the advantages of the two is the OPPRF-PSI (also dubbed `circuit-based PSI') which utilizes the advances in OPRF as well as providing interface for subsequent analytics based on MPC   \cite{pinkas2019efficient,GMR+21}.

However, challenges arise as requirement for privacy and efficiency is escalated. Firstly, in the situation that data owners are allowed to know the shared IDs, the renowned PSI protocols suffice to meet the requirement \cite{RS21,RR22}. However, when more stringent privacy requirement is imposed such that one or even both of the data owners are not allowed to learn the shared IDs, these PSI protocols are no longer sufficient. We need to resort to the framework of generic MPC or OPPRF-PSI. Secondly, a plaintext inner join returns linked records associated with shared IDs exclusively while its secure realizations usually not. For instance, OPPRF-PSI returns secret shares of inner join as well as some redundancies \cite{pinkas2019efficient}. It is challenging to remove the redundancies if shared IDs are to be protected. The reason is straightforward: a participant will only remove a record if he is aware that it does not belong to the overlap of the two datasets. This contradicts the privacy requirement that the shared IDs between the datasets should not be disclosed. Oblivious shuffling was used to mitigate this issue \cite{JSZ+22,MSZ15,CGP+20,GMR+21}.
An example is in \cite{GMR+21} where oblivious switching (i.e., an amortized shuffling) is used twice to remove the unmatched records obliviously. Nevertheless, oblivious shuffling breaks the linkage between IDs and their associated attributes. As a result, it only supports attributes from one side while attributes from the other side are missing. Thirdly, there exists a solution for secure inner join where shared IDs are also protected in the project of LANCELOT \cite{LANCELOT2021,innerjoin}. However, the number of public key operations scales linearly with the overall size of the dataset, making it impractical to handle large datasets in use cases like advertising and banking where there are even billions of records in the datasets.


\subsection{Contribution}
In this work, we developed a technique, namely PSA, to efficiently and securely inner join two datasets of different stewardships. We defined PSA to achieve two levels of privacy protection, i.e., \textit{single blinded} and \textit{double blinded}, depending on the availability of the shared IDs. The motivation for this classification is obtained from two types of use cases. Firstly, there are circumstances where one data owner is allowed to learn the shared IDs while the other data owner is not. A typical example is privacy-preserving Graph Convolution where one party claims unlabeled topology while the other party has labeled nodes. The one knowing the topology will be in a better position to run the convolution if he learns the common nodes, whereas existing studies essentially unveil shared IDs to both parties \cite{cheung2021fedsgc}. Secondly, \textit{double blinded} PSA is devised for the use cases (e.g., advertising conversion) when final analytics on the inner join is to be disclosed rather than any intermediate result amid the process of PSA.

The contributions of this work are summarized as follows.

Firstly, we developed the PSA protocols for the above two privacy levels. The proposed protocols  provide interface to subsequent MPC operations, facilitating further analytics tasks. At the core of PSA is a modified oblivious switching (OSN) network which leverages symmetric key operations and offline pre-computation to save online run time.

Secondly, we implemented and benchmarked the performance of the proposed protocols in different network conditions. To give an example, on a medium fast network (500 Mbps bandwidth), the proposed PSA protocol computes inner join of two datasets at the scale of one million records in 35.5 sec on a single thread. In comparison, its counterpart, the HE-based solution, accomplishes in 56.3 mins with the same setup, indicating an approximately $\times 100$ speedup by PSA.

Thirdly, we gave a new proof for the existence of a quasi-linear algorithm to construct an oblivious switching network to achieve arbitrary permutation, as is distinct from the existing ones in the literature \cite{GMR+21}.

In e-commerce, we sometimes need to process datasets at the scale of billions of records. To address this issue, our modified OSN allows us to shift most of the sub-procedures offline, leaving only a small portion to be handled online. This design is particularly beneficial for use cases involving datasets at the scale of billions of records as will be discussed in Section~\ref{sec:offline}.

\section{Related Studies}
Existing works to facilitate privacy-preserving inner join can be classified into two predominant methodological classes. The first class encompasses the MPC-enabled oblivious query systems that employ garbled-circuit and secret-sharing to find the matches and return the linked records without revealing the sensitive data \cite{bater2017smcql, volgushev2019conclave,liagouris2021secrecy}. The other class is characterized by the utilization of PSI protocols and their variants to find the matches in a privacy-preserving manner \cite{reichert2021circuit,ion2020deploying,pinkas2019efficient}.

Oblivious query systems, exemplified by Conclave\cite{volgushev2019conclave}, SMCQL\cite{bater2017smcql}, Senate\cite{poddar2021senate}, etc., represent a crucial advancement in oblivious queries for statistical data extraction from databases. 
A noteworthy progression in this domain is the introduction of the Secrecy \cite{liagouris2021secrecy}, which represents a paradigm shift towards prioritizing privacy preservation as it was designed to target the situation that data owners do not have enough computation resources to finish the collaborative analytics, and they need to outsource the data by replicated secret sharing. However, the reliance on MPC for all operations inherently introduces latency and computational complexities.

The other line of research follows the advances in PSI. As a high level description, common IDs are found by a PSI protocol before the associated attributes are selected and reciprocally shared with the other participants. Designs of this kind are showcased by Alipay in their SCQL project \cite{secretflow23} and by AWS in their Clean Rooms project. However, when the computation parties are not allowed to know the common IDs, to blindly select the attributes associated with the common ID is challenging. 

Another work of the second class is the Private Computation on Set Intersection (PCSI) based on Oblivious Switching Network (OSN), underscoring the utility of OSN in confidential computation over the intersection without revealing it \cite{GMR+21}. Subsequent studies on OSN and its application to confidential computation can be found in \cite{CGP+20,JSZ+22}. However, these advancements are not without their limitations. With OSN for two-party PCSI \cite{GMR+21}, the attributes from only 1 party is blindly selected for computation.

A concurrent work for privacy-preserving inner join with a semi-honest helper is the one in LANCELOT project \cite{LANCELOT2021} for heart failure prediction. This server-aided framework supports attributes from both sides and it makes PSI even simpler compared with a standard two-party PSI. However, the required Paillier-based public key operations increase linearly with the total number of records in the two datasets, making it inefficient for large-scale datasets.


\section{Preliminaries}\label{sec:preli}
\subsection{Notations}
We denote by $[n]$ the set $\{0,1,\cdots,n-1\}$. We use $(A_i)_{i\in[n]}$ as a shorthand for a vector $\vec{A}=(A_0,\cdots,A_{n-1})$, and we denote by $A_i$ the element at coordinate $i$. We omit the header $\vec{}$ for a vector (e.g., $A=(A_i)_{i\in[n]}$) when there is no uncertainty in the context. For an array $M$ of size $m\times n$, we use $M_{i,j}$ to denote an element at column $i$ and row $j$. We also use $M_a[i][j]$ alternatively when the array is annotated by a subscript $_a$. This convention also applies to vectors for indexing. We denote by $M_{i,:}$ (\textit{resp.} $M_{:,j}$) a column vector (\textit{resp.} row vector) of array $M$. We use operator $:=$ to assign value(s), and we use $==$ to compare if two operands are equal. We use $\pi:\mathcal{D}\rightarrow\mathcal{C}$ to denote a mapping from a set $\mathcal{D}$ to a set $\mathcal{C}$. We denote by $\llbracket x \rrbracket = (\llbracket x \rrbracket_a, \llbracket x \rrbracket_b)$ the replicated secret sharing of $x$ where $\llbracket x \rrbracket_a$ and $\llbracket x \rrbracket_b$ are two shares. We denote by $x\leftarrow_R \mathcal{D}$ sampling $x$ from domain $\mathcal{D}$ uniformly at random. 
When an operator (e.g. secret sharing or permutation) applies to a vector operand $A=(A_i)_{i\in[n]}$, it applies  coordinate-wise to every element $A_i$. We denote by $\log$ for base 2 logarithm. We use $\oplus$ for bit-wise exclusive or, and denote by $a\|b$ the concatenation of 2 strings. We denote by $\pi_1 \cdot \pi_2$ sequential execution of permutation $\pi_1$ and $\pi_2$. We use the caligraphic font $\mathcal{F}_a$ for the ideal functionality of $a$ and use $\Pi_a$ to represent the protocol to achieve the functionality. 
\subsection{Private Set Alignment}
The functionality of private set alignment (PSA) is given in Figure.~\ref{algo:funinnerjoin}. Let $(X, U)$ and $(Y, V)$ be the two datasets from two participants, with $X,Y$ to be the identifiers and $U,V$ to be the attributes. For every $z\in X\cap Y$, we denote by $u_z$ and $v_z$ the two associated attributes with identifier $z$. 

The ideal PSA takes the datasets from two participants and returns the secret shares of the combined and aligned attributes denoted by $\llbracket (u_z,v_z)  \rrbracket$ for every $z\in X \cap Y$. It is a primary requirement that neither party sees the other's dataset in clear. Moreover, we define two levels of privacy protection depending on the availability of the intersection $X\cap Y$. A \textit{single blinded} PSA allows party $P_1$ to learn $X\cap Y$ and $u_z$ while the other party only learns $|X\cap Y|$. For a \textit{double blinded} PSA, no parties learn $X\cap Y$ and $(u_z,v_z)$ but $|X\cap Y|$.   

\begin{figure}
    \centering
\begin{mdframed}
\textit{Setting:} Assume $P_1$ and $P_2$ each have a structured and relational dataset sharing the same identifier system but with different attributes. To ease the notation, we assign 1 attribute for each dataset.

\textit{Inputs:}
    \begin{itemize}
        \item Participant $P_1$ inputs the dataset denoted by pairs of ID and attribute as $(X, U)= \{(x_1, u_{x_1}),  (x_2, u_{x_2}), \ldots, (x_n, u_{x_n})\}$.
        \item Participant $P_2$ also inputs the dataset denoted by pairs of ID and attribute as $(Y,V) = \{(y_1, v_{y_1}), (y_2,v_{y_2}), \ldots, (y_m,v_{y_m}) \}$. 
    \end{itemize}
    \textit{Outputs:}
    \begin{itemize}
        \item Both participants receive the secret shares of the inner join as $\llbracket (u_{z}, v_{z}) \rrbracket$ for every $z\in X \cap Y$ where $u_{z}$ and $v_{z}$ are associated payload.
    \end{itemize}

    \textit{Privacy Guarantee:}
    \begin{itemize}
        \item \textit{level 1 (single blinded)} $P_1$ learns nothing but $X \cap Y$ and $u_{z}$. $P_2$ learns nothing but the cardinality $|X\cap Y|$;
        \item \textit{level 2 (double blinded)} $P_1$ and $P_2$ learns nothing but the cardinality $|X\cap Y|$. 
    \end{itemize}
\end{mdframed}
    
    \caption{Ideal functionality of private set alignment between two datasets}
    \label{algo:funinnerjoin}
\end{figure}

\subsection{Private Set Intersection}
Private Set Intersection (PSI) is a customized form of multiparty computation that enables two or more parties, each possessing a private set of items, to identify the common items in their sets without revealing the items beyond the intersection. PSI is generalized to provide versatile functionalities to meet different use cases and privacy requirements. The ideal functionality of a standard two-party PSI is given in Figure~\ref{fig:psi}. To satisfy privacy protection of level 1 as defined in Figure~\ref{algo:funinnerjoin}, we will be using a PSI protocol based on vector oblivious linear evaluation (VOLE) and oblivious key-value storage (OKVS) from \cite{RR22}; for level 2 privay, we will employ a server-aided PSI where the two parties only learn the cardinality of the intersection.

\begin{figure}
    \centering
\begin{mdframed}
    \textit{Inputs:}
    \begin{itemize}
        \item A receiver inputs elements $X = \{x_1, x_2, \ldots, x_n\}$;
        \item A sender inputs elements $Y = \{y_1, y_2, \ldots, y_m\}$.
    \end{itemize}

    \textit{Outputs:}
    \begin{itemize}
        \item Receiver outputs $ X \cap Y$;
        \item Sender outputs nothing.
    \end{itemize}
\end{mdframed}

    \caption{Ideal functionality of PSI.}
    \label{fig:psi}
\end{figure}
\subsection{Oblivious Switch Network} \label{sec:osnpreli}
The notion of oblivious switching network was proposed in \cite{MS13} as one of the pillars of  circuit topology hiding for private function evaluation. The ideal functionality of OSN is given in Figure.~\ref{fig:funcOSN}. The sender of OSN inputs a vector $X$ and the receiver inputs a permutation $\pi$ on his choice. As a result, the two parties output the secret shares of the permuted $X$ with respect to $\pi$. It is oblivious in the sense that neither party learns the other's input.
It is notable that the mapping $\pi:[N]\rightarrow[M]$ can be bijection, injection, and surjection. In this work, we use bijection and injection for PSA because the inner joined table has no more rows than input tables. Realization of OSN is based on the technique of rearrangeable network (e.g. Bene\v{s} network) and its oblivious evaluation. The underlying idea is any mapping $\pi:[N]\rightarrow[M]$ can be realized in quasi-linear complexity by a rearrangeable network which consists of some fixed structures as well as some programmable structures. For the fixed ones, they are publicly known. For the programmable ones, (a) the OSN receiver can determines those structures with quasi-linear complexity according to the mapping $\pi$ (b) the two parties of OSN can obliviously evaluate the rearrangeable network such that the receiver does not learn vector $X$ while the sender does not learn the programmable structures as well as $\pi$. Taking Bene\v{s} network as an example, we provide the details about rearrangeable network and its oblivious evaluation in Section \ref{sec:proof} as well as in Appendix.~\ref{app:struBens}, \ref{app:osnroutines} and \ref{app:evalBenes}. 

\begin{figure}
    \centering
    \begin{mdframed}
        \textit{Parameters:}
        \begin{itemize} 
            \item mapping $\pi:[N]\rightarrow[M]$ for arbitrary $N,M\in \mathbb{Z}^+$; some parameter $l\in\mathbb{Z}^+$.
        \end{itemize}
        \textit{Inputs:}
        \begin{itemize}
            \item Sender inputs a vector $X = (x_0, x_1, \ldots, x_{M-1})$ where $x_i\in\{0,1\}^l$.
            \item Receiver inputs a permutation $\pi:[N]\rightarrow[M]$.
        \end{itemize}
        \textit{Outputs:}
        \begin{itemize}
            \item Sender and receiver output the secret shares of the permuted vector as 
            $\llbracket\tilde{X}\rrbracket = (\llbracket x_{\pi(0)} \rrbracket, \llbracket x_{\pi(1)} \rrbracket, \cdots, \llbracket x_{\pi(N-1)} \rrbracket)$.
        \end{itemize}
    \end{mdframed}
    \caption{Ideal functionality of OSN.}
    \label{fig:funcOSN}
\end{figure}
\section{A new Proof: to programme Bene\v{s} Network with quasi-linear complexity }\label{sec:proof}
The realization of OSN is accomplished in two steps. At the first step the sender programmes a rearrangeable network according to a desired permutation $\pi$. Then, the sender and receiver obliviously evaluate the network such that they output secret shares of the permuted data as in Figure~\ref{fig:funcOSN}. In this section, we give a new proof for the complexity of the programming step.

\begin{figure}
    \centering
    \includegraphics[scale=0.8]{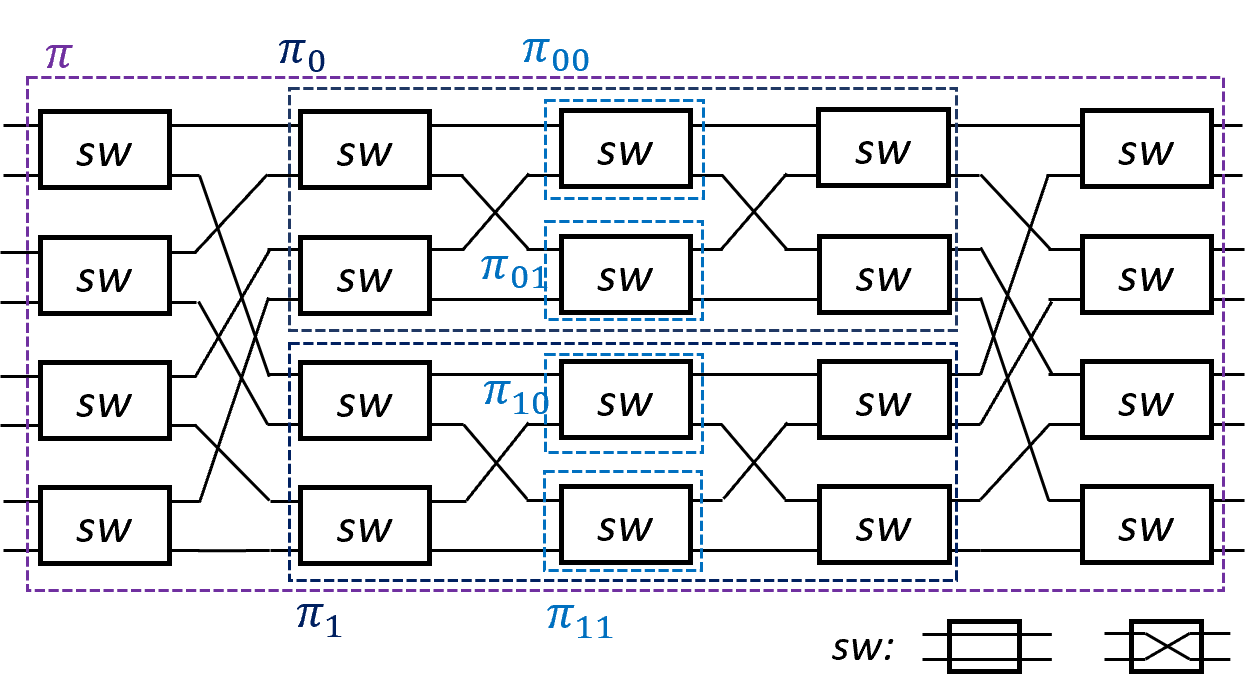}
    \caption{Construction of Bene\v{s} network: an example for $N=8$, the permutation $\pi$ is recursively divided into sub-permutations $\pi_b,\pi_{bb}$ for binary $b$.}
    \label{fig:senderBenes}
\end{figure}

An example of Bene\v{s} network, an instance of rearrangeable network, to achieve arbitrary 8-by-8 bijection is depicted in Figure~\ref{fig:senderBenes}. Bene\v{s} network was initially designed to achieve $N$-by-$N$ bijection for a 2-power $N$. It is built with an array of base switch gates and the ``hard-coded'' wires connecting the base switch gates. The array has $\frac{N}{2}$ rows and $2\log {N} -1$ columns. 
The base switch gates, each having two incoming wires and two outgoing wires, are programmed to be either a ``crossover'' or a ``straight-through'' according to the permutation $\pi$ to achieve.  The solution to programme the network for a target permutation is not unique. The process of defining the base switch gates is called the \textit{looping} algorithm which is elucidated in Appendix~\ref{app:osnroutines}. It recursively divides an $N$-by-$N$ permutation $\pi$ into two $\frac{N}{2}$-by-$\frac{N}{2}$ permutations $\pi_0$ and $\pi_1$, which are progressively divided into another two permutations $\pi_{00}, \pi_{01}$ and $\pi_{10}, \pi_{11}$. The process stops until a $2$-by-$2$ bijection is reached.

 The \textit{looping} algorithm provides a systematic way to programme the network according to a permutation $\pi$ in quasi-linear complexity. We in this section give an interpretation and proves the complexity of such an algorithm in a different manner from that in \cite{GMR+21}. 


\begin{figure}
    \centering
    \begin{subfigure}[b]{0.2\textwidth}
        \centering
        \includegraphics[width=\textwidth]{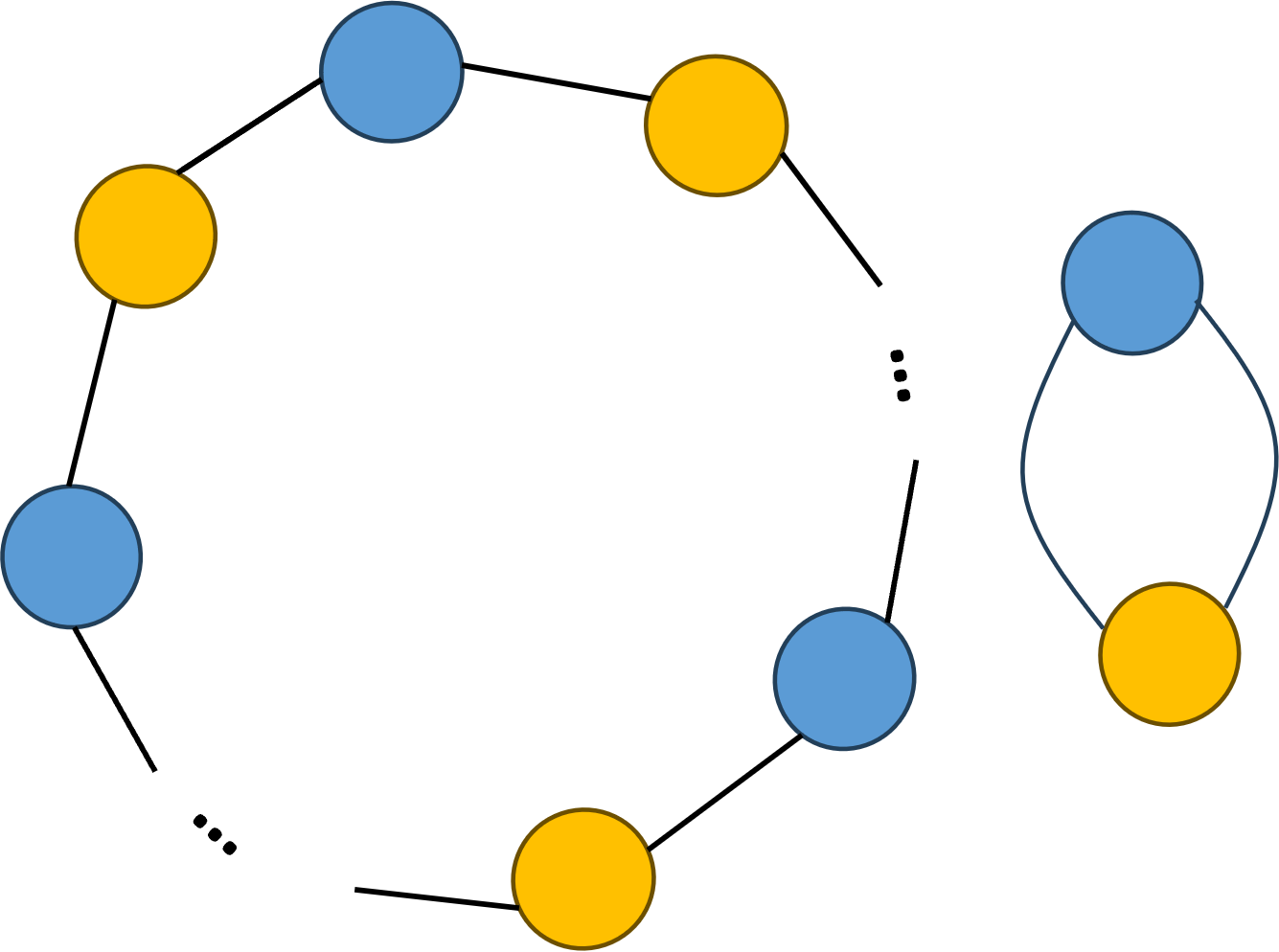}
        \caption{Two-power $N$: even cycles.}
        \label{fig:graphsubfig1}
    \end{subfigure}
    ~
    \begin{subfigure}[b]{0.2\textwidth}
        \centering
        \includegraphics[width=\textwidth]{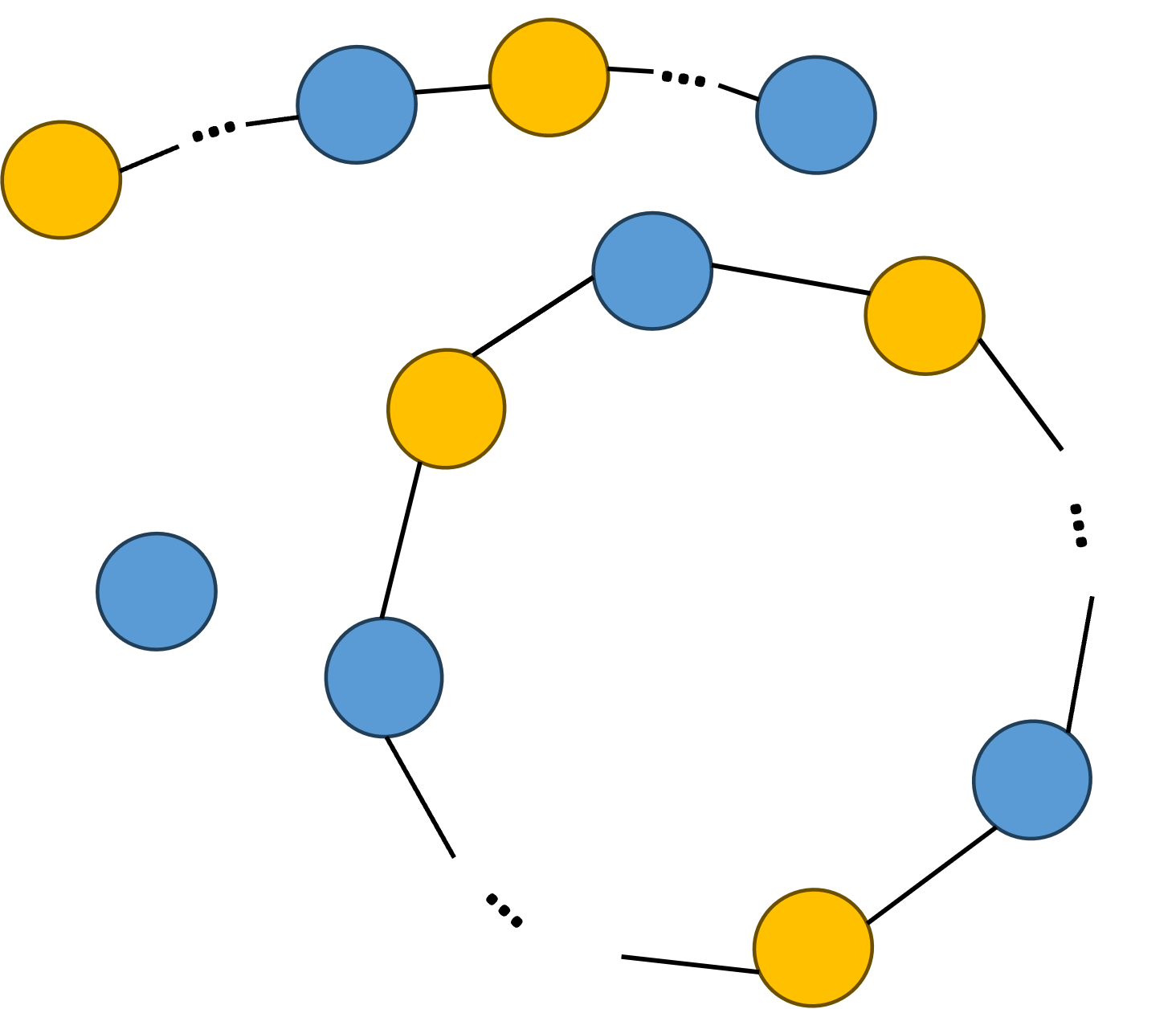}
        \caption{Arbitrary $N$: even cycles, chain and node.}
        \label{fig:graphsubfig2}
    \end{subfigure}
    \caption{Graph coloring for a graph defined by permutation $\pi:[N]\rightarrow[N]$.}
    \label{fig:graph}
\end{figure}

\begin{definition}[\cite{OT70}]\label{def:dual}
We group a set $[N]$ for some 2-power $N$ into $N/2$ subgroups such that each subgroup is denoted by
\[
J(l,2) := \{i \mid \lfloor i/2 \rfloor= l \},
\]
for $l=0,1,\cdots,\lfloor \frac{N-1}{2}\rfloor$. We call the 2 items in the same subgroup the \textit{dual} of each other, and $l$ is termed the \textit{characteristic} of the subgroup $J(l,2)$.
\end{definition}
\begin{lemma}\label{lem:2powerN}
    There exists an algorithm to programme a Bene\v{s} network at quasi-linear complexity $O(N \log {N})$ to achieve any permutation denoted by $\pi:[N]\rightarrow[N]$ for any $2$-power $N$. 
\end{lemma} 
\begin{proof}
Denote by $\pi(0),\pi(1),\cdots,\pi({N-1})$ the \textit{permutation} of sequence $0,1,\cdots,N-1$. We also define a \textit{inverse permutation} $\pi^{-1}(0),\pi^{-1}(1),\cdots,\pi^{-1}({N-1})$ such that $\pi(\pi^{-1}(i))=i$ for $i=0,1,\cdots,N-1$. 

    When $N=2^k$ for some $k\in\mathbb{Z}^+$, we define a undirected graph with a vertex set as
    \[
    V:= \left\{0,1,\cdots,N-1 \right\}.
    \]

Given the vertices $V$ and a permutation $\pi:[N]\rightarrow[N]$, the edges of this undirected graph is defined as follows. We connect the vertices $i$ and $j$ if either one of the two situations is met. (a) The vertices $i$ and $j$ are dual of each other (i.e. $i,j\in J(l,2)$ for some $l\in[N/2-1]$) (b) If $\pi^{-1}(i)$ and $\pi^{-1}(j)$ are dual of each other (i.e. $\pi^{-1}(i),\pi^{-1}(j)\in J(l,2)$ for some $l\in[N/2-1]$). The reason why we define the edges in this way is two-folded. Firstly, at the input gate layer (leftmost column) of Bene\v{s} network (e.g., in Figure~\ref{fig:senderBenes}), the process of dividing a permutation equally into 2 sub-permutations is done by selecting one incoming wire from each input gate (i.e. select 1 item in each subgroup $J(l,2)$ for every $l\in[N/2-1]$) and assigning it to one sub-permutation (upper sub network) while assigning the other wire (i.e. their dual items) to the other sub-permutation (lower sub network). Secondly, at the output gate layer (rightmost column) of the network, the 2 outgoing wires of any output gate (the 2 items in any subgroup $J(l,2)$) are connected to either the upper or the lower sub network, exclusively. In a nutshell, the edges are defined to reflect the constraint imposed by ``hard-coded'' wires between input/output-layer  and the layers in between.
    
   Because vertex $i$ and its dual will be assigned to two different sub-permutations and so are vertex $\pi^{-1}(i)$ and its dual. The problem of determining an gate at input layer to be a ``straight-through'' or a ``corssover'' is equivalent to a \textit{graph coloring problem}
    (to paint the color of each vertex to be ``0'' or ``1'') under the constraint that adjacent vertices must be in different colors. In other words, we aim to prove this graph is \textit{bipartite}.
    This can be justified by showing that this graph comprises of \textit{simple even cycles} exclusively. We can derive this conclusion immediately 
    because each vertex has a degree exactly equal to 2 and every vertex and its dual must reside in the same cycle. An graphical interpretation is in Figure~\ref{fig:graphsubfig1}. As a result, we can conclude that the graph is bipartite and therefore the \textit{graph coloring problem} can be solved in complexity $O(N)$ by a \textit{deep first search}. 
    
    Once the graph coloring problem is done, we know how to programme all the gates at input and output layer accordingly. Then we recursively proceed to the two sub-permutations and define the input and output layers in the same fashion until we reach a $2$-by-$2$ permutation. 
    For 2-power $N$, the size of each sub-permutation is always even and the same proof will hold. Because the recursion depth is $\log {N}$, we can conclude there exists a solution to define every gate of the Bene\v{s} network in $O(N\log {N})$ complexity.
\end{proof}

Bene\v{s} network was later developed to achieve bijection and even injection and surjection for arbitrary $N$. Without loss of generality, we in this work will use the terminology generalized Bene\v{s} network to indicate all its derivatives. A detailed construction of generalized Bene\v{s} network is in Appendix~\ref{app:struBens} and an example for $9$-by-$9$ bijection is given in Figure~\ref{fig:genrlBenes}. We derive the quasi-linear algorithm to programme the generalized Bene\v{s} network as a corollary of Lemma~\ref{lem:2powerN}. The \textit{looping} algorithm is elucidated as Algorithm \ref{Algo:looping} in Appendix \ref{app:osnroutines}.

\begin{corollary}\label{lem:arbitraryN}
There exists an algorithm to programme a generalized Bene\v{s} network at quasi-linear complexity $O(N\log N)$ to any permutation $\pi:[N]\rightarrow[N]$ for arbitrary $N$.
\end{corollary}
\begin{proof}
    This proof will inherit the definition of graph and the convention from Lemma~\ref{lem:2powerN}. Additionally, if N is odd, the last item itself $N-1$ will constitute a subgroup $J(l,2)$ for $l=\lceil N/2 \rceil$. 
    
    To prove a $N$-by-$N$ permutation can be divided into two sub-permutations subject to the constraint of the ``hard-coded'' wires is equivalent to the \textit{graph coloring problem} as in Lemma~\ref{lem:2powerN}. 
    However, as opposed to a graph comprised of even cycles exclusively for $2$-power $N$, the graph for arbitrary $N\in\mathbb{Z}^+$ may comprise of \textit{even cycles} together with a \textit{node} or a \textit{chain} as in Figure~\ref{fig:graphsubfig2}. Again, the cycles must be even because every vertex has degree 2 and it resides on the same cycle as its dual. It is evident that a node will occur if the last item $N-1$ is not permuted. On the contrary, a permuted vertex $N-1$ will give rise to a chain because it has degree one. No matter for \textit{simple even cycles} cycles, chain, or node, the graph is bipartite hence the \textit{graph coloring problem} is solvable by \textit{deep first search} with complexity $O(N)$. According to the results of graph coloring, we programme all the gates at input and output layer such that we derive two sub-permutations. 

    Note that for odd $N$ as in Figure~\ref{fig:genrlBenes}, the `hard-coded' wires split a permutation into a sub-permutation of size $\lfloor N/2 \rfloor$ and the other one of size $\lceil N/2 \rceil$. The recursion will terminate until a $2$-by-$2$ or $3$-by-$3$ bijection is reached (i.e. the sub-network in the middle in Figure~\ref{fig:genrlBenes}). The same proof will hold recursively and the recursion depth is $\lfloor \log {N} \rfloor$. As a result, there exists an algorithm of complexity $O(N\log N)$ to programme the generalized Bene\v{s} network.
    
\end{proof}

Bene\v{s} network is generalized to achieve an injection $\pi:[M]\rightarrow[N]$ for some $M\le N$ \cite{GMR+21}. In a nutshell, they prune some base switch gates such that it achieves an injective mapping from output to input. It reduces the complexity of OSN to $O(N\log M)$; nonetheless, the enhancement from $N\log N$ to $N\log M$ is not substantial especially for vertically partitioned datasets where the overlap is large.
Moreover, a disadvantage of this method is that we are allowed to programme the network only after $M$ is known. For PSA, $M$ is known online when PSI is done. To shift the programming step to an offline phase, we proposed an easier way to achieve injection from Bene\v{s} network, as illustrated in Figure~\ref{proto:mosn} of the Section \ref{sec:lvl1}. 

\section{Problem Statement and A Framework} \label{sec:ps}
\subsection{Problem Statement and Challenges}
\begin{figure}
    \centering
    \includegraphics[scale=0.35]{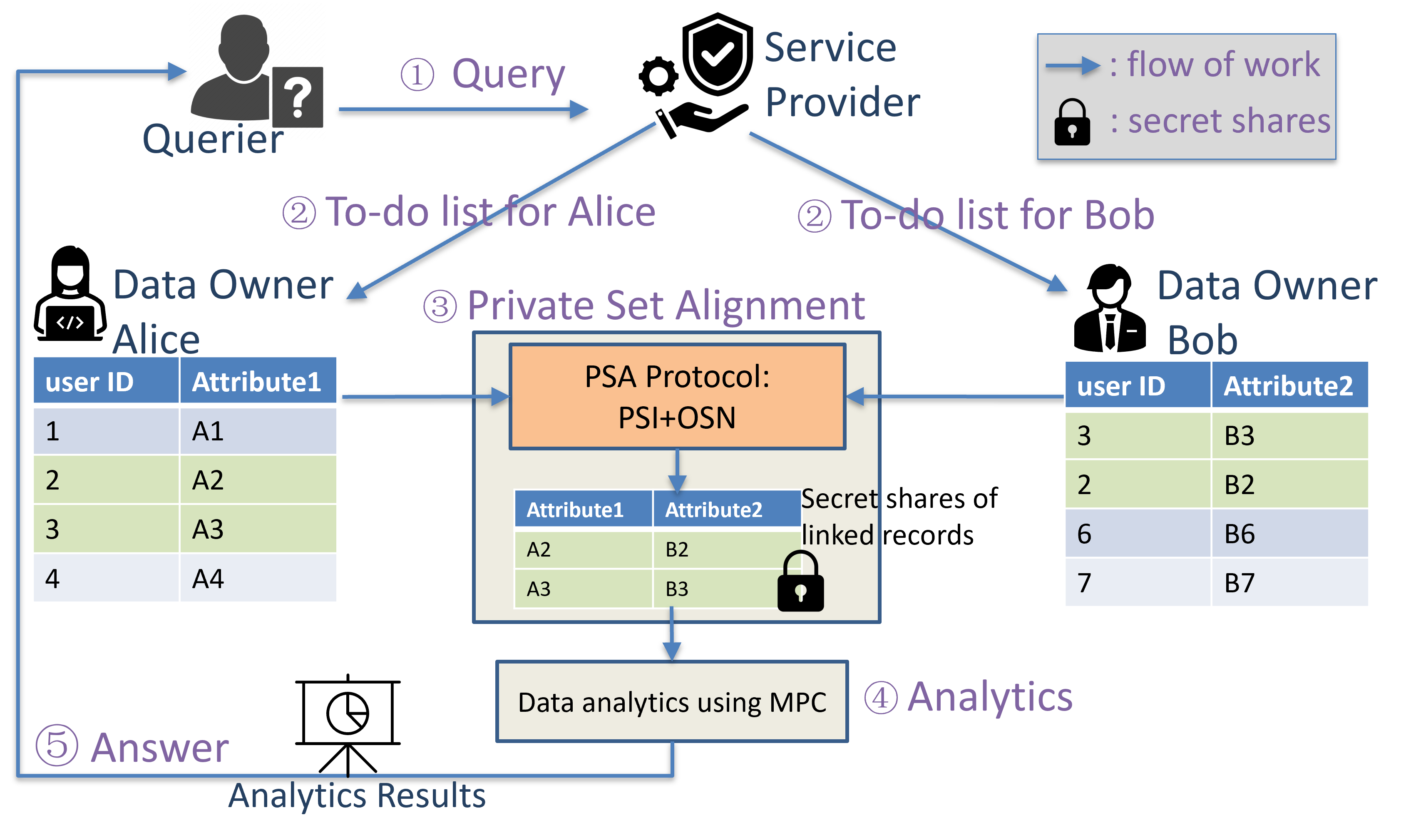}
    \caption{A flow of work of secure and collaborative query on vertically partitioned data.}
    \label{fig:obliQuery}
\end{figure}

The research problem is motivated by the use case of Secure Collaborative Query Language (SCQL), as is a part of Alipay's open-source project \cite{secretflow23}, where a querier expects oblivious analytics on two vertically partitioned datasets of different stewardship. Ahead of the analytics is the private set alignment (PSA) by which the two parties get secret shares of a ``virtual table'' as the inner join of the two datasets with respect to common user IDs. A flow of work of this use case is depicted in Figure~\ref{fig:obliQuery}. Description of each participant is given as below.
\begin{itemize}
    \item Querier: query the service provider for certain analytics on distributed datasets.
    \item Service provider: generate to-do lists for data owners to give the analytics result. 
    \item Data owners: carry out a sequence of secure and collaborative operations according to the to-do list; answer the querier with the analytics result. Additionally, a data owner can also play a dual role of querier in certain circumstances.
\end{itemize}
The research problem we deal in this work is to develop an efficient PSA protocol to meet the privacy requirements as formalized in Figure~\ref{algo:funinnerjoin} of Section \ref{sec:preli}. It is worth noting that the challenge substantially arises at \textit{privacy level 2}, the \textit{double blinded} PSA. The reasons are threefold. 
Firstly, it is challenging for both data owners to blindly remove unmatched records while matched ones are aligned. Secondly, by the existing serverless method in \cite{GMR+21}, the two data owners can blindly remove the unmatched records and create a ``virtual table'' which, however, consists of attribute from one data owner only. If we want to have the selected records associated with common IDs from the other party in the ``virtual table'', there is not evident solution. Thirdly, a server-aided solution showcases poor computation and communication efficiency, particularly for large datasets commonly seen in e-commerce use cases, due to the use of somewhat homomorphic encryption \cite{LANCELOT2020,innerjoin}.
\subsection{Our solution and assumptions for private set alignment}\label{sec:assumptions}
In this work, we propose our solution (i.e. Figure.~\ref{fig:obliQuery}) for private set alignment as integration of PSI and oblivious switching network. Protocols to meet the privacy requirements will be given in the next section. Especially, to achieve \textit{privacy level 2} as defined in Figure.~\ref{algo:funinnerjoin}, we propose an efficient server-aided private set alignment by leveraging the engagement of service provider who normally only takes charge of compilation and coordination in oblivious queries systems \cite{secretflow23}. Moreover, this solution provides interface to generic MPC-based analytics on the linked records.

 We assume all participants are semi-honest and non-colluding. The fundamental trust in the service provider is established by laws and regulation and is viewed at the cornerstone of the business of this type. The service provider will risk of business failure and even worse if collusion is revealed to compromise the privacy of any data owners. 
\begin{figure}
    \centering
\begin{mdframed}
\textbf{Inputs:}
    \begin{itemize}
        \item Participant $P_1$ inputs the dataset denoted by pairs of ID and attribute as  $(X, U)= \{(x_0, u_{0}),  (x_1, u_{1}), \ldots, (x_{n-1}, u_{n-1})\}$.
        \item Participant $P_2$ also inputs the dataset denoted by pairs of ID and attribute as $(Y,V) = \{(y_0, v_{0}), (y_1,v_{1}), \ldots, (y_{m-1},v_{m-1}) \}$.
    \end{itemize}
\textbf{Output:}
Participant $P_1$ and $P_2$ learn the secret shares of inner join $((\llbracket u_{\pi_1(i)}\rrbracket,\llbracket v_{\pi_2(i)}\rrbracket))_{i\in[c]}$. Additionally, $P_1$ learns $X\cap Y$ and $P_2$ learns $|X\cap Y|$.

    \textbf{Procedure:}
    \begin{itemize}
        \item[1.] $P_1$ and $P_2$ invoke an OPRF protocol $\Pi_{oprf}$ with their IDs as input; as a result, $P_1$ derives a vector of PRF values
        \[
        \tilde{X}:=(\textsf{PRF}(x_i))_i, i=0,\cdots,n-1,
        \]
        and $P_2$ also derives a vector of PRF values 
        \[
        \tilde{Y}:= (\textsf{PRF}(y_i))_i, i=0,\cdots,m-1,
        \]
        
        \item[2.] $P_2$ sends  $\tilde{Y}$ to $P_1$.
        \item[3.] Let $c=|\tilde{X}\cap\tilde{Y}|$. $P_1$ defines a vector $J:=(J_i)_{i\in[c]}$  such that $\tilde{Y}_{J_i}\in\tilde{X}\cap\tilde{Y}$ for $J_i\in[m]$. Then $P_1$ shuffles the coordinates of $J$ randomly. To ease the notation, we denote by $J$ the shuffled vector, as well. $P_1$ also defines the other vector $K:=(K_i)_{i\in[c]}$ such that $K_i\in[n]$ and $\tilde{X}_{K_i} = \tilde{Y}_{J_i}$ for every $i\in[c]$. The two vectors $J$ and $K$ consist of the indices  of elements in $\tilde{X}\cap\tilde{Y}$ \textit{w.r.t.} vector $\tilde{Y}$ and $\tilde{X}$, respectively, giving rise to two injections $\pi_1:[c]\rightarrow [m]$ and $\pi_2:[c]\rightarrow [n]$.
        
        \item[4.] $P_1$ and $P_2$ invoke the modified OSN protocol $\Pi_{mosn}$ with $P_1$ inputting the injection $\pi_1$ and $P_2$ inputting vector $V=(v_{0},v_{1},\cdots,v_{m-1})$. As a result, $P_1$ and $P_2$ learns $\llbracket v_{{\pi_1(0)}} \rrbracket, \llbracket v_{{\pi_1(1)}} \rrbracket,\cdots, \llbracket v_{{\pi_1(c-1)}} \rrbracket $.
        \item[5.] $P_1$ selects $u_{{\pi_2(0)}},u_{{\pi_2(1)}},\cdots, u_{{\pi_2(c-1)}}$. For each $u_{\pi_2(i)}$, $P_1$ splits it into 2 replicated shares and sends $\llbracket u_{\pi_2(i)}\rrbracket_2$  to $P_2$.
    \end{itemize}

\end{mdframed}
    \caption{Private set alignment protocol for \textit{privacy level 1}.}
    \label{proto:joinlvl1}
\end{figure}
\section{Our Protocols}
In this section, we give the proposed PSA protocols for the two privacy levels as defined in Figure~\ref{algo:funinnerjoin}. We use oblivious pseudorandom function (OPRF), a modified oblivious switching network (OSN) as sub-protocols denoted by $\Pi_{oprf}$ and $\Pi_{mosn}$, respectively.
\begin{figure}
\centering
\begin{mdframed}
\textbf{Parameters:}
 A base field $\mathbb{B}$ and its extension $\mathbb{F}$; random oracles $H^\mathbb{B}: \{0,1\}^* \rightarrow \mathbb{B}$, $H^o: \mathbb{F} \rightarrow \{0,1\}^{\text{out}}$;
 
\textbf{Input:}
 ${P}_1$: $X = \{x_0, \ldots, x_{n-1}\}$; ${P}_2:~\emptyset$
 
\textbf{Output:}
    ${P}_1$: ${X}' = \{\textsf{PRF}(x_i)\mid x_i \in X, i=0,\cdots,n-1\}$; ${P}_2$: the seed to evaluate a \textsf{PRF} with elements of his choice.

\textbf{Procedure:}
\begin{itemize}
    \item[1.] ${P}_1$ samples a randomness $r$;
    \item[2.] ${P}_1$ defines a set of key-value pairs $L = \{(x_i, H^\mathbb{B}(x_i)) \mid x_i \in X, i=0,\cdots,n-1\}$ and computes $\vec{P}$ = \textsf{okvs.Encode}$(L, r)\in\mathbb{F}^m$ for some $m\ge n$;
    \item[3.] ${P}_1$ and ${P}_2$ invoke a $\mathcal{F}_{vole}$ with dimension $m$. As a result, ${P}_2$ outputs $\vec{B}\in\mathbb{F}^m$ and $\Delta\in\mathbb{F}$, and ${P}_1$ learns $\vec{C}\in\mathbb{F}^m$, $\vec{A}\in\mathbb{B}^m$ s.t. $\vec{C} = \vec{A}\Delta + \vec{B}$;
    \item[4.] ${P}_1$ sends $r$ and $\vec{A}'$ to ${P}_2$ where $\vec{A}' = \vec{A} + \vec{P}$
    \item[5.] ${P}_2$ derives $\vec{B}' = \vec{B} + \vec{A}'\Delta$ as a seed to evaluate a \textsf{PRF} as 
     $H^o(\textsf{okvs.Decode}(\vec{B}', y, r) - H^{\mathbb{B}}(y)\Delta)$ for any $y$ of his choice.
     \item[6.] ${P}_1$ outputs a set of \textsf{PRF} values ${X}' = \{\textsf{PRF}(x_i):=H^o(\textsf{okvs.Decode}(\vec{C}, x_i, r)) \mid x_i\in \vec{X}\}$;

\end{itemize}
\end{mdframed}
\caption{An OPRF protocol $\Pi_{oprf}$ based on OKVS and VOLE \cite{RR22,GPR+21}}\label{proto:oprf}
\end{figure}

\begin{figure}[ht]
\centering
\begin{mdframed}
\textbf{Input:}
 ${P}_1$: $\pi:[c]\rightarrow[m]$ for some $c\leq m$; ${P}_2$: $U = (u_0, \ldots, u_{n-1})$ for $u_i\in\{0,1\}^*$.
 
\textbf{Output:}
\item $P_1$ and $P_2$ receive the secret shares of the permuted vector i.e. $\llbracket \tilde{U}\rrbracket = \{\llbracket u_{\pi(0)} \rrbracket, \llbracket u_{\pi(1)} \rrbracket, \ldots, \llbracket u_{\pi(c-1)}\rrbracket\}$;

\textbf{Procedure:}
\begin{itemize}
    \item[1.] (\textit{offline}) $P_1$ generates a random $m$-to-$m$ bijection denoted by $\rho_1:[m]\rightarrow[m]$. Then, $P_1$ runs the subroutine $looping(\lceil \log m\rceil,0, m-1, 0, [m], \rho_1([m]))$ as in Algorithm~\ref{Algo:looping} in Appendix~\ref{app:osnroutines} such that a  Bene\v{s} network of size $(2\lceil \log m \rceil -1)\times m$ is programmed to achieve the permutation $\rho_1:[m]\rightarrow[m]$.
    \item[2.] $P_2$ declares two $(2\lceil \log m \rceil-1)$-by-$m$ arrays denoted by $A$ and $B$. As depicted in Figure~\ref{fig:senderOSNeval} of Appendix~\ref{app:evalBenes}, $P_2$ assigns random values $A_{i,j}$ and $B_{i,j}$ to every incoming and outgoing wire of the base switch gates of Bene\v{s} network, respectively.
    If there is a ``hardcoded'' connection between outgoing wire $j_1$ of column $i-1$ and incoming wire $j_2$ of column $i$, $P_2$ will assign $A_{i,j_2}=B_{i-1,j_1}$. For simplicity of description, we omit $A_{i,j}$ for $i\neq 0$ in Figure~\ref{fig:senderOSNeval}.
    \item[3.] $P_2$ sends $P_1$ a vector of masked values as $\tilde{U}=(u_j\oplus A_{0,j})_{j\in[m]}$. 
    \item[4.] $P_1$ and $P_2$ invoke OT extension to generate one 1-out-2 OT instance for every base switch gate. For each OT instance, $P_2$ inputs $A_{i,j_0} \oplus B_{i,j_0}\| A_{i,j_1} \oplus B_{i,j_1} $ and $A_{i,j_0} \oplus B_{i,j_1}\| A_{i,j_1} \oplus B_{i,j_0}$. On the other side, $P_1$ inputs his selection of either 0 (``straight-through") or 1 (``crossover") as derived from $looping$ algorithm in step 1.
    \item[5.] Given all the OT output, $P_1$ invokes the subroutine $evaluate(m,0, 0, \tilde{U})$ as in Algorithm~\ref{Algo:receiverEval} in Appendix~\ref{app:osnroutines} and derives vector $(u_{\rho_1(j)}\oplus B_{R,j} )_{j\in[m]}$ where $R=2\lceil \log m \rceil -2$ indexes the last column of array $B$.
    \item[6.] $P_1$ deduces the inverse $\rho_1^{-1}:[m]\rightarrow[m]$ of $\rho_1$. $P_1$ computes an injection $\rho_2=\pi \cdot \rho_1^{-1}$ as $\rho_2(i)=\rho_1(\pi(i))$ for $i\in[c]$, and sends $\rho_2$ to $P_2$. As a result, $P_1$ outputs $\llbracket u_{\pi(i)}\rrbracket_1:= u_{\rho_1(\rho_2(i))}\oplus B_{R,\rho_2(i)}$ for $i\in[c]$.
    \item[7.] Given $\rho_2$, $P_2$ indexes the column vector $B_{R,:}$ of array $B$ by $\rho_2$ and outputs $\llbracket u_{\pi(i)}\rrbracket_2:=B_{R,\rho_2(i)}$ for $i\in[c]$. 

\end{itemize}
\end{mdframed}
\caption{A modified Oblivious Switching Network protocol $\Pi_{mosn}$ \cite{MS13,JSZ+22,GMR+21}}\label{proto:mosn}
\end{figure}
\subsection{Protocol for privacy level 1}\label{sec:lvl1}
For \textit{privacy level 1}, one data owner is allowed to learn the common IDs. So he is able to explicitly select the associated attributes and arranges them in certain order on his own choice. The other data owner learns nothing beyond the the number of common IDs. We provide the private set alignment protocol and its sub-protocols to achieve \textit{privacy level 1} in Figure~\ref{proto:joinlvl1}, \ref{proto:oprf}, \ref{proto:mosn}. 

At step 2 in Figure~\ref{proto:joinlvl1}, as in most PSI protocols, $P_2$ is supposed to shuffle the PRF vector $\tilde{Y}$ before it is sent to $P_1$ to prevent leaking the positions of common IDs. At step 4, the same shuffling also applies to the attribute vector $V$ such that the modified OSN protocol $\Pi_{mosn}$ returns the attributes associated with common IDs correctly. We omit this step only for simplicity of notations. Another consideration to randomize the location of records in the ``virtual table" is step~3 where $P_1$ randomly shuffles $J$. This operation will also implicitly shuffle vector $K$. As a result, the ``virtual table" $(\llbracket u_{\pi_2(i)} \rrbracket, \llbracket v_{\pi_1(i)} \rrbracket)_{i\in[c]}$ will be arranged with the rows shuffled. Otherwise, the joined records will be arranged in the same sequence as in $P_1$'s input dataset.

Besides, there are a few choices for the OPRF protocol $\Pi_{oprf}$. We provide an instance based on \textit{oblivious key value storage} (OKVS) \cite{GPR+21} and \textit{vector oblivious linear evaluation} (VOLE) \cite{RR22} in Figure~\ref{proto:oprf}, which gives state-of-the-art performance for the purpose of computing the intersection in a two-party situation. At step 2 of $\Pi_{oprf}$ in Figure~\ref{proto:oprf}, we denote by $\mathcal{F}_{vole}$ the ideal functionality of VOLE. In addition, \textsf{okvs.Encode}$(L,r)$ is an encoding algorithm taking the key-value pairs $L$ and some randomness $r$ as input and yielding a vector $\vec{P}\in\mathbb{F}^m$ as output. At step 5 and step 6, the \textsf{okvs.Decode}$(\cdot)$ is a decoding algorithm. It takes as input a vector over $\mathbb{F}^m$, the same randomness $r$ as in step 1, and arbitrary elements from the same domain as $X$ and $Y$, and it yields a value over $\mathbb{F}$ as output.


One of the contributions of this work is the modified OSN protocol which shifts some procedures of PSA to an offline phase as in Figure~\ref{proto:mosn}. The modified OSN protocol $\Pi_{mson}$ is derived from the standard OSN whose functionality and realization are discussed a lot in Section \ref{sec:osnpreli}, \ref{sec:proof} and Appendix \ref{app:struBens}, \ref{app:osnroutines}, \ref{app:evalBenes}. 
In the context of private set alignment, we expect OSN to achieve an injection $\pi:[c]\rightarrow[m]$ where $c=|X\cap Y|$. Due to the fact that $c$ is determined upon the accomplishment of PSI, $P_1$ is only able to programme the injection network online. In this work, as described in Figure~\ref{proto:mosn}, we instead let $P_1$, the OSN receiver, programme a bijection network offline for arbitrary $\rho_1:[m]\rightarrow[m]$ on his own choice. In other words, $looping$ (Algorithm~\ref{Algo:looping}, Appendix~\ref{app:osnroutines}) of complexity $O(m\log m)$ is performed offline at step 1 of $\Pi_{mons}$ to save the run time in online phase. From step 2 to step 5 in Figure~\ref{proto:mosn}, $P_1$ and $P_2$ obliviously evaluate the bijection network with respect to $\rho_1$ such that $P_1$ and $P_2$ derive a shuffled and secret shared $U$ as $\llbracket u_{\rho_1(i)}\rrbracket$ for $i\in[m]$. $P_1$ knows the target injection $\pi:[c]\rightarrow[m]$
and $\rho_1$, so he can immediately derive an intermediate injection $\rho_2:=\pi \cdot \rho_1^{-1}$ and sends $\rho_2$ to $P_2$. As $P_2$ does not know $\pi$ and $\rho_1$, he can only infer $\pi$ from $\rho_2$ no better than random guessing. Eventually, $P_1$ and $P_2$ index $\llbracket u_{\rho_1(i)}\rrbracket$ with $\rho_2$ and gets $(\llbracket u_{\pi(i)}\rrbracket)_i$ because of the relationship $\pi=\rho_2\cdot \rho_1$.

\begin{figure}[!ht]
    \centering
\begin{mdframed}
\textbf{Parameters:} a random oracle $H:\{0,1\}^{\kappa_1}\times\{0,1\}^*\rightarrow\{0,1\}^{\kappa_2}$;

\textbf{Inputs:}
    \begin{itemize}
        \item $P_1$: a dataset denoted by pairs of ID and attribute as  $(X, U)= \{(x_0, u_{}),  (x_1, u_{1}), \ldots, (x_{n-1}, u_{n-1})\}$.
        \item $P_2$: a dataset denoted by pairs of ID and attribute as $(Y,V) = \{(y_0, v_{0}), (y_1,v_{1}), \ldots, (y_{m-1},v_{m-1}) \}$.
        \item \textit{server}: $\emptyset$.
    \end{itemize}
\textbf{Output:}
Participants $P_1$ and $P_2$ learn secret shares of inner join $((\llbracket u_{\pi_1(i)}\rrbracket,\llbracket v_{\pi_2(i)}\rrbracket))_{i\in[c]}$. Additionally, \textit{server} learns the cardinality $c:=|X\cap Y|$ of the intersection as well as $n$ and $m$.

    \textbf{Procedure:}
    \begin{itemize}
        \item[1.] 
        $P_1$ and $P_2$ agree on a key $r\in\{0,1\}^{\kappa_1}$ by any key agreement protocol.
        \item[2.] $P_1$ evaluates $H(r,\cdot)$ on input $X=(x_i)_{i\in[n]}$ and gets a vector of PRF values as
        \[
        \tilde{X}:=(\textsf{PRF}(x_i))_i, i=0,\cdots,n-1.
        \]
        Then, $P_1$ sends $\tilde{X}$ to \textit{server} via a secure channel. 
        \item[3.] $P_2$ evaluates $H(r,\cdot)$ on input $Y=(y_i)_{i\in[m]}$ and gets a vector of PRF values as
        \[
        \tilde{Y}:= (\textsf{PRF}(y_i))_i, i=0,\cdots,m-1.
        \]
        Then, $P_2$ sends $\tilde{Y}$ to \textit{server} via a secure channel. 
        
        \item[4.] Upon receiving $\tilde{X}$ and $\tilde{Y}$, \textit{server} computes $\tilde{X}\cap\tilde{Y}$. Let $c=|\tilde{X}\cap\tilde{Y}|$.
        \item[5.] \textit{Server} defines a vector $J:=(J_i)_{i\in[c]}$  such that $\tilde{Y}_{J_i}\in\tilde{X}\cap\tilde{Y}$ for $J_i\in[m]$. He shuffles the coordinates of $J$ randomly.
         For simplicity, we denote the shuffled vector by $J$, as well. \textit{Server} defines the other vector $K:=(K_i)_{i\in[c]}$ such that $\tilde{X}_{K_i} = \tilde{Y}_{J_i}$ and $K_i\in[n]$ for every $i\in[c]$. The index vector $J$ and $K$ immediately induce two injections $\pi_2:[c]\rightarrow [m]$ and $\pi_1:[c]\rightarrow [n]$, respectively.
        
        \item[6.] \textit{Server} and $P_1$ invoke the modified OSN protocol $\Pi_{mosn}$ with \textit{server} inputting the injection $\pi_1$ and $P_1$ inputting vector $U=(u_{0},u_{1},\cdots,u_{n-1})$. As a result, \textit{server} and $P_1$ learns $\llbracket u_{{\pi_1(0)}} \rrbracket,\llbracket u_{{\pi_1(1)}} \rrbracket,\cdots, \llbracket v_{{\pi_1(c-1)}} \rrbracket $.
        \item[7.] \textit{Server} sends $P_2$ his $\Pi_{mosn}$ output $\llbracket u_{{\pi_1(0)}} \rrbracket_{2},\llbracket u_{{\pi_1(1)}} \rrbracket_{2},\cdots, \llbracket v_{{\pi_1(c-1)}} \rrbracket_{2}$ via a secure channel.
        \item[8.] \textit{Server} and $P_2$ repeat the procedure of step 6 and 7 with \textit{server} inputting the injection $\pi_2$ and $P_2$ inputting $V=(v_i)_{i\in[m]}$. As a result, $P_1$ and $P_2$ get secret shares of inner join $(\llbracket u_{{\pi_1(i)}} \rrbracket, \llbracket v_{{\pi_2(i)}} \rrbracket)_{i\in[c]}$. 
    \end{itemize}

\end{mdframed}
    \caption{Private set alignment protocol for \textit{privacy level 2}.}
    \label{proto:joinlvl2}
\end{figure}
\subsection{Protocol for privacy level 2}\label{sec:lvl2}
Incorporating a semi-honest third-party server into various protocols can notably enhance computing power, reliability, and functionality. For instance, Abadi et al.\cite{abadi2022multi} introduced a delegated semi-honest computation server for an advanced multi-party Private Set Intersection protocol, overcoming the limitations of traditional protocols. Similarly, other studies\cite{duong2020catalic}\cite{kavousi2020improved}\cite{debnath2021secure} have leveraged computing servers to boost computational capacity for devices with limited power, ensuring privacy protection of identifiers. TNO's Lancelot project\cite{innerjoin} also adopted this strategy for secure inner join, employing a semi-honest server to filter out unmatched records while safeguarding ID privacy by a cryptographic hash with its seed shared between data owners. However, scalability is limited by the computational and communication cost associated with Paillier cryptosystem.

As discussed in previous sections, if neither of the two data owners is allowed to know the common IDs amid the process of the private set alignment, it is challenging to establish the record linkage between two datasets. To achieve \textit{privacy level 2} is for the two parties to obliviously select the attributes associated with the common IDs and arrange them in the same order all the while the two parties have no idea about what they select and in what order they are arranged. In the use case of SCQL \cite{secretflow23} for secure and collaborative data analytics as in Figure.~\ref{fig:obliQuery}, there is a server as the service provider and more importantly a coordinator and compliler to translate queries from client into privacy-preserving operations for data owners. This server is assumed to be semi-honest as stated in Section~\ref{sec:assumptions}. We in this section leverage the semi-honest server to achieve \textit{privacy level 2} for private set alignment. The proposed protocol is described in Figure.~\ref{proto:joinlvl2}.

We employ an AES-based hash in Matyas-Meyer-Oseas (MMO) mode to realize the random oracle $H:\{0,1\}^{\kappa_1}\times\{0,1\}^*\rightarrow\{0,1\}^{\kappa_2}$ for $\kappa_1=\kappa_2=128$. According to the birthday attack paradox, it would take approximately $2.6\times 10^{13}$ attempts by brute force to find the 1st collision with probability at least $2^{-40}$ if the output of $H$ is equally probable. In e-commerce use cases, the datasets under consideration are often in the millions or billions. Therefore, the correctness of the protocol is ensured because unmatched inputs will not give identical $\textsf{PRF}$ values almost surely. Moreover, from a security perspective, AES-128 suffices to provide a computational complexity of $2^{117}$ against a pseudo preimage attack \cite{BDG+2020,CGL+24}. So, the semi-honest server is computationally incapable to compute the preimage of $H$.

At step~2 and step 3, $P_1$ and $P_2$ shuffle their \textsf{PRF} vectors before they send them to \textit{server}, meanwhile they shuffle their datasets simultaneously. We omit this step only for simplicity of description. At step~5 and step 6, \textit{server} computes the injection $\pi_1$ and $\pi_2$ according to $\tilde{X}\cap\tilde{Y}$ and shuffles them simultaneously such that the ``virtual table'' $(\llbracket u_{\pi_1(i)}\rrbracket,\llbracket v_{\pi_2(i)} \rrbracket)_{i\in[c]}$ is arranged with the rows shuffled randomly.

In this PSA protocol for \textit{privacy level 2}, \textit{server} learns the cardinality $c:=|X\cap Y|$ of the intersection as well as $n$ and $m$. We consider this information leakage to be acceptable in the use case of SCQL as in Figure~\ref{fig:obliQuery} as $c$ itself doesn't unveil raw data and the parties are non-colluding as stated in Section~\ref{sec:assumptions}.

\section{Implementation and Performance}
This section details the performance evaluation of our proposed private set alignment protocol with a comparison with an existing method utilizing somewhat homomorphic encryption \cite{innerjoin}. The benchmarking was conducted in a controlled laboratory environment on a Dell Precision 7856 workstation equipped with CPU of AMD Ryzen PRO @ 3.8GHz and RAM of 128GB. The performance was evaluated on a single thread.

We firstly provide the standalone benchmarks of sub-procedures as PSI and OSN. Additionally, we utilize the Linux Traffic Control command to simulate various bandwidth settings to evaluate the overall performance of the protocol.

\subsection{Standalone Benchmarks}
The protocol we proposed in Figure. \ref{proto:joinlvl2} comprises of two sub-procedures: step 1 through 5 implement a server-aided private set intersection, while by step 6 through 8 invoke two instances of modified OSN. To isolate the computational efficiency, we benchmarked the performance the PSI and modified OSN without bandwidth influence in the standalone tests. The results are summarized in Table \ref{tab:stdaloneserverpsi} and \ref{tab:stdaloneOSN} where dataset size is the number of records in each dataset and we assume the two datasets are equally large without loss of generality.  

It is observed from the two tables that the modified OSN dominates the overall computation and communication complexity. Moreover, we also showcase the run time of both online and offline phase, from which we observe the offline \textit{looping} algorithm to be more time consuming than the online phase particularly for large datasets. 
For the use case of SCQL (Figure~\ref{fig:obliQuery}) with a third-party server as service provider, the server is able to accomplish the offline procedure by itself in idle time as long as it knows the size of the datasets. 
\begin{table*}[!h]
    \centering
    \caption{Standalone performance for the server-aided PSI.}
    \begin{tabular}{|l|*6{p{1cm}|}}
        \hline
         \multicolumn{1}{|l|}{PSI} & \multicolumn{6}{c|}{ Standalone test without bandwidth influence} \\
        \hline
        dataset size & $2^{12}$ & $2^{14}$ & $2^{16}$ & $2^{18}$ & $2^{20}$ & $2^{22}$\\
        \hline
        time (ms) & 0.251 & 0.549 & 1.94 & 24.38 & 103.71 & 491.12 \\
        \hline
        communication (MB) & 0.13 & 0.50 & 2.00 & 8.00 & 32.00 & 128.00  \\
        \hline
    \end{tabular}
    \label{tab:stdaloneserverpsi}
\end{table*}

\begin{table*}[htbp]
    \centering
    \caption{Standalone performance for the modified OSN.}
    \begin{tabular}{|l|*6{p{1cm}|}}
        \hline
         \multicolumn{1}{|l|}{modified OSN} & \multicolumn{6}{c|}{ standalone test  without bandwidth influence} \\
        \hline
        dataset size & $2^{10}$ & $2^{12}$ & $2^{14}$ & $2^{16}$ & $2^{18}$ & $2^{20}$\\
        \hline
        time: online phase (ms) & 30.3 & 32.9 & 51.8 & 150.1 & 591.8 & 2501.4 \\
        \hline
        time: offline phase (ms) & 0.8 & 2.8 & 7.1 &46.0  & 469.1 & 7876.3 \\
\hline
        communication (MB) & 0.47 & 2.33 & 10.38 & 47.51 & 214.01 & 952.01  \\
        \hline
    \end{tabular}
    \label{tab:stdaloneOSN}
\end{table*}

\subsection{Benchmarks with Bandwidth Settings}
For a realistic assessment, we evaluate the performance of the PSA protocol as in Figure~\ref{proto:joinlvl2} in various bandwidth settings. In our implementation, we use a computational security strength of 128 bits.
Without loss of generality, we assume two equally large datasets with half of the records having common IDs.
The performance is given in Table \ref{tab:TCinnerjoin}.
It is observed that the proposed protocol gives remarkable run time performance in high bandwidth environment while it scales down when the bandwidth is limited. This is due to the fact that the communication overhead of modified OSN results in a transmission delay when bandwidth is low. Nonetheless, the proposed protocol outperforms the HE-based one substantially.


\begin{table*}[htbp]
    \centering
    \caption{Traffic-control benchmarks for PSA of privacy level 2 : our method vs. the HE-based one. Run time in seconds and communication in MB.}
    \begin{tabular}{|l|*{15}{p{0.60cm}|}}
        \hline
        \multicolumn{1}{|l|}{\makecell{ \\[1pt] \textbf{Protocol}}}  & \multicolumn{3}{c|}{\textbf{10 Gbits/sec (s)}}   & \multicolumn{3}{c|}{\textbf{1 Gbits/sec (s)}} & \multicolumn{3}{c|}{ \textbf{500 Mbits/sec (s)}} & \multicolumn{3}{c|}{ \textbf{200 Mbits/sec (s)}} &  \multicolumn{3}{c|}{\textbf{Comm. (MB)}}    \\
        \makecell{(dataset size)}  & $2^{16}$ & $2^{18}$ & $2^{20}$ & $2^{16}$ & $2^{18}$ & $2^{20}$ & $2^{16}$ & $2^{18}$ & $2^{20}$ & $2^{16}$ & $2^{18}$ & $2^{20}$ & $2^{16}$ & $2^{18}$ & $2^{20}$\\
        \hline
        \makecell{ Our protocol\\[2pt]}  & 0.2 & 0.9 & 4.9 & 0.9 & 4.2 & 19.2 & 1.7 & 7.8 & 35.5 & 4.1 & 18.7 & 84.8 & 99.0  & 444.0 & 1968.0  \\
        \hline
        \makecell{ HE-based protocol \\  \cite{innerjoin}}  & 211.8  & 840.2 & 3553.8 & 212.5  & 842.9 & 3364.6 & 213.3 & 846.0 &3377.1  & 215.6 & 855.3 & 3414.0 &  96.0 & 384.0 &  1536.0  \\
        \hline
    \end{tabular}
    \label{tab:TCinnerjoin}
\end{table*} 

\begin{table}[htbp]
    \centering
    \caption{Standalone benchmarks for Paillier Cryptosystem with key size $=3072$ bits at security strength of 128 bits.}
    \begin{tabular}{|c|c|}
        \hline
         \textbf{Operations}  & {\makecell{\textbf{Millisecond per operation}}} \\
        \hline
        key generation & 1175.16\\
        \hline
        encryption  & 0.25 \\
        \hline
        decryption  & 2.38 \\
        \hline
        addition  & 0.00841\\
        \hline
        subtraction & 0.065 \\
        \hline
    \end{tabular}
    \label{tab:HE}
\end{table}

\subsection{Comparative Study}
We also provide the performance of the HE-based inner join in \cite{innerjoin} which also employs a third-party server to achieve the same functionality as PSA for \textit{privacy level 2}. The comparison is shown in Table~\ref{tab:TCinnerjoin}. Their project is implemented by python available at \url{https://github.com/TNO-MPC/protocols.secure_inner_join}, while we implemented our protocol with C++. To ensure a fair comparison, we benchmarked the Paillier cryptosystem as in Table~\ref{tab:HE} \footnote{Available at https://github.com/abb-iss/ophelib} in our experimental environment. As the complexity of the HE-based inner join scales linearly with the number of encryption and decryption operations, we can immediately estimate its run time and communication overhead. In the estimation, we assume the same environment, single thread configuration, and the same security level as those used in our implementation. They transmission delay is approximated with the highest transmission speed in each bandwidth setting, giving it an advantage over our implementation. The detailed calculations are in Appendix~\ref{app:estimate}.


As shown in Table~\ref{tab:TCinnerjoin}, if we assume a medium fast network (e.g., 500 Mbps), PSA is approximately 100 times faster than the HE-based inner join for two datasets of $2^{20}$ records each.
We can conclude that our protocol substantially outperforms the HE-based approach to achieve the same level of privacy protection, providing an more effective solution particularly for large datasets.

\subsection{Shifting More Offline For Dataset of Billions with Records}\label{sec:offline}
In e-commerce sector, enterprises are stewarding data of billions of customers. This poses a challenges to PSA. If we assume a 1 Gbps bandwidth and scale up the run time linearly, to accomplish PSA for two one-billion datasets will take 2.7 hours\footnote{To get a realistic estimation, we no longer use the single thread setup and assume two data owners run modified OSN with server simultaneously.}. The run time is substantial, but existing methods seem no better. However, the proposed modified OSN shows an advantage that the bulk of the work can be done offline leaving only minor work to be done online. 

To this end, we shift all the operations and communications not involving business data to an offline phase. In Figure~\ref{proto:mosn}, those operations occur in step 1, 2 and 4, where two parties are working on random data independent of raw business data. Other steps, which involve sending and processing business data (e.g., $u_i$ and $\pi$), are executed online. In this way, the majority of procedures from step 6 through step 8 in Figure~\ref{proto:joinlvl2} are done offline. Only the server-aided PSI (step 1 through step 5 in Figure~\ref{proto:joinlvl2}), Algorithm \ref{Algo:receiverEval} \textit{evaluate}($\cdot$) and computing $\rho_2 = \pi\cdot \rho_2^{-1}$ are performed online, making it orders of magnitude easier than the offline tasks.

According to Table~\ref{tab:stdaloneserverpsi}, server-aided PSI is done efficiently. Algorithm~\ref{Algo:receiverEval} is also efficient whose performance is given in Table~\ref{tab:maskevaluate} in Appendix~\ref{app:evalBenes}. Besides, computing $\rho_2$ can be done easily by an ordered map container. In summary, this approach makes it feasible to accomplish PSA for datasets at the scale of billions of records within minutes.

\section{Security Proof}
In this section, we formalize the security proof of our protocols, focusing on their resistance against semi-honest and non-colluding adversaries. We assume adversaries possess polynomial-time computing resources. We analyze the protocols within the real/ideal simulation paradigm.

\begin{definition}
    A protocol $\Pi$ is semi-honest secure with respect to ideal functionality $\mathcal{F}$ with input $X$ from party $P_0$ and input $Y$ from party $P_1$ if there exists probabilistic polynomial time (PPT) simulator $\textsf{Sim}_0$ and $\textsf{Sim}_1$ such that a PPT adversary who passively corrupted a party is not able to distinguish the simulator's output and $\mathcal{F}(X,Y)$ from the corrupted party's view and protocol output. This notion is described as follows.
    \begin{align*}
        &(\textsf{view}_0^{\Pi}(X, Y), \textsf{Out}_0^{\Pi}(X,Y)) \stackrel{c}{\approx} (\textsf{Sim}_0(X,\mathcal{F}(X,Y)), \mathcal{F}(X,Y)), \\
        &(\textsf{view}_1^{\Pi}(X, Y), \textsf{Out}_1^{\Pi}(X,Y)) \stackrel{c}{\approx} (\textsf{Sim}_1(Y,\mathcal{F}(X,Y)), \mathcal{F}(X,Y)).
    \end{align*}
\end{definition}
As we see, the proposed private set alignment protocols are composed of sub protocols including $\Pi_{oprf}$ and $\Pi_{mosn}$. The former is secure under our security assumptions which is proved in the literature \cite{RR22,GPR+21}. The original OSN is also proved to be secure in \cite{GMR+21,JSZ+22}. We developed our modified OSN based on an OSN for bijection $\rho_1$. To transform $\rho_1$ into a target injection $\pi$, the OSN sender computes $\rho_2  = \pi \cdot \rho_1^{-1}$ where $\pi,\rho_1$ is known by OSN sender exclusively while $\rho_2$ is known by both sides. As discussed in Section~\ref{sec:lvl1}, given $\rho_2$ the OSN sender does not show advantage in learning about $\pi$ over random guessing. A simulation-based proof for $\Pi_{mosn}$ is give in Theorem~\ref{thm:osnproof}.
\begin{theorem} Given a secure OT extension protocol $\Pi_{OText}$, the modified OSN protocol $\Pi_{mosn}$ in Figure.~\ref{proto:mosn} is secure against semi-honest adversaries. \label{thm:osnproof}
\end{theorem}
\begin{proof} The simulation proceeds as follows.

 a) $P_2$ is passively corrupted.
    
\textit{Simulation}. For any PPT adversary $\mathcal{A}_2$ controlling $P_2$ in the real world, we describe a simulator $\textsf{Sim}_2$ who simulates $\mathcal{A}_2$'s view in the ideal world. The revisions to $\Pi_{mosn}$ are as follows.
    \begin{itemize}
        \item[1)] As in step 2 - 3 in Figure.~\ref{proto:mosn}, $\mathcal{A}_2$ declares random matrix A and B and sends the masked input $\tilde{U}$ to $\textsf{Sim}_2$. 
        \item[2)] As in step 4 in Figure.~\ref{proto:mosn}, $\mathcal{A}_2$ computes $A_{i,j_0} \oplus B_{i,j_0}\| A_{i,j_1} \oplus B_{i,j_1} $ and $A_{i,j_0} \oplus B_{i,j_1}\| A_{i,j_1} \oplus B_{i,j_0}$ for each base switch gate as his input to the 1-out-2 OT.
        \item[3)] At step 1, $\textsf{Sim}_2$ randomly selects a bijection $\rho_1':[m]\rightarrow[m]$ and programme the array of switches using the $looping(\cdot)$ algorithm. As a result, $\textsf{Sim}_2$ derives his selecting bits for every 1-out-2 OT instances for the network.
        \item[4)] At step 5, upon seeing the $\mathcal{A}_2$'s input to OTs and the masked vector $\tilde{U}$, $\textsf{Sim}_2$ evaluates the Bene\v{s} network programmed according to $\rho_1'$ and evaluate the network.
        \item[5)] At step 6, $\textsf{Sim}_2$ also randomly select a injection $\pi':[c]\rightarrow[m]$ and sends $\rho_2'=\pi' \cdot \rho_1'^{-1}$ to $\mathcal{A}_2$.
        \item[6)] Finally, $\mathcal{A}_2$ output $B_{m-1,\rho_2'(i)}$ for $i\in [c]$. This ends the description of simulator $\textsf{Sim}_2$.
    \end{itemize} 

 \textit{Indistinguishability}. Under the assumption of a secure OT extension protocol, $\mathcal{A}_2$ receives only 1 message i.e. $\rho_2'$, as the simulation of $\rho_2$ in real world. Since $\rho_1'$ and $\rho_1$ are randomly selected $m$-to-$m$ bijections and are identically distributed, so are $\rho_2'$ and $\rho_2$. Besides, $\mathcal{A}_2$'s output $B_{R,\rho_2'(i)}$ for $i\in[c]$ is also randomly distributed as its counterpart in real world. So, $\mathcal{A}_2$ is not able to distinguish the simulation from real world. 

b) $P_1$ is passively corrupted.

    \textit{Simulation}. For any PPT adversary $\mathcal{A}_1$ that controls $P_1$ in the real world, we describe a simulator $\textsf{Sim}_1$ who simulates $\mathcal{A}_1$'s view in ideal world. The revisions to $\Pi_{mosn}$ are as follows.
    \begin{itemize}
        \item[1)] At step 2 in Figure~\ref{proto:mosn}, $\textsf{Sim}_1$ generates random values for every incoming and outgoing wires on the Bene\v{s} network. 
        \item[2)] At step 3, $\textsf{Sim}_1$ randomly generates $P_2$'s input vector $U'=(u'_i)_{i\in[m]}$ and masks it with the random values assigned to incoming wires at input layer of Bene\v{s} network as $\tilde{U}'=(\tilde{u}'_i)_{i\in[m]}$.
        \item[3)] At step 4, $\textsf{Sim}_1$ computes $A_{i,j_0} \oplus B_{i,j_0}\| A_{i,j_1} \oplus B_{i,j_1} $ or $A_{i,j_0} \oplus B_{i,j_1}\| A_{i,j_1} \oplus B_{i,j_0}$ created on his own choice and inputs them to the 1-out-2 OT instances for the network. $\mathcal{A}_1$ inputs his selecting bits to 1-out-2 OTs. 
        \item[4)] At step 5 and 6, upon receiving $\tilde{U}'$ and the 1-out-2 OT output for every switch gate, $\mathcal{A}_1$ evaluates the network on the received message. $\mathcal{A}_1$ also computes $\rho_2=\pi \cdot \rho_1^{-1}$ and sends it to $\textsf{Sim}_1$.
        \item[5)] Finally, $\mathcal{A}_1$ outputs $B_{m-1,\rho_2(i)}\oplus u'_{\rho_1(\rho_2(i))}$ for $i\in[c]$.
    \end{itemize}
 
 \textit{Indistinguishability}. Under the assumption of a secure OT extension protocol, $\mathcal{A}_1$ receives 2 messages i.e. the 1-out-2 OT output on the selecting bits of his choice for every switch gate, and the masked vector $\tilde{U}'$. They has identical distribution with their genuine counterparts in real world, and so does $\mathcal{A}_1$'s output $u'_{\rho_1(\rho_2(i))}\oplus B_{R,\rho_2(i)}$ for $i\in[c]$. As a result, a PPT $\mathcal{A}_1$ is not able to distinguish the simulation from real world protocol.
 
\end{proof}

\begin{theorem}[Protocol for \textit{privacy level 1}]
The protocol in Figure.~\ref{proto:joinlvl1} realizes ideal functionality $\mathcal{F}_{psa}$ at \textit{privacy level 1} in semi-honest setting.
\end{theorem}
\begin{proof}
Given the security guarantee provided by $\Pi_{oprf}$ and $\Pi_{mosn}$, we give the proof sketch as follows.

a) In case $P_1$ is passively corrupted, the messages received by $P_1$ are the PRF vectors $\tilde{X},\tilde{Y}$ at step 1-2 and the the secret shares of $P_2$'s attribute $V$ subject to injection $\pi_1$ at step 5 in Figure~\ref{proto:joinlvl1}. Besides, $P_1$ outputs $X\cap Y$.

The simulator $\textsf{Sim}_1$ learns $P_1$'s input and learns $X\cap Y$ from ideal functionality $\mathcal{F}_{psa}$ for \textit{privacy level 1}. 
Then $\textsf{Sim}_1$ generates a random vector $\tilde{X}'=(\tilde{X}_i')_{i\in[n]}$
and sends it to $P_1$. Additionally, $\textsf{Sim}_1$ defines $\mathcal{I}:=\{i\in[c]\mid c=|X\cap Y|, x_i \in X\cap Y \}$. Then $\textsf{Sim}_1$ defines and shuffles the vector $\tilde{Y}':= ( (\tilde{X}'_i)_{i\in \mathcal{I}} \| z_1, \cdots, z_{m-c} )$ where $z_j$ are random dummies and sends it to $P_1$. $P_1$ is not able to distinguish between $\tilde{X}'$ and $\tilde{X}$ assuming a secure $\Pi_{oprf}$; the same conclusion holds for $\tilde{Y}'$. Moreover, $\tilde{Y}'$ is defined such that $\tilde{X}'\cap\tilde{Y}'$ indicates the same common IDs as $\tilde{X}\cap\tilde{Y}$ does. Besides, the output of $\Pi_{mosn}$ is also random and indistinguishable from that of an ideal $\mathcal{F}_{mosn}$.

b) In the case $P_2$ is corrupted, it is evident that the indistinguishability holds because there are no additional messages added to $P_2$'s view  other than output of $\Pi_{oprf}$ and $\Pi_{mosn}$.

\end{proof}

\begin{theorem}[Protocol for \textit{privacy level 2}]
    The protocol in Figure.~\ref{proto:joinlvl2} realizes ideal functionality $\mathcal{F}_{psa}$ at \textit{privacy level 2} in a semi-honest and non-colluding setting.
\end{theorem}
\begin{proof}
The security of PSA for \textit{privacy level 2} is ensured by the secure key agreement, the AES-128 hash function $H:\{0,1\}^{\kappa}\times\{0,1\}^\ast\rightarrow\{0,1\}^{\kappa}$, and $\Pi_{mosn}$.

a) Either $P_1$ or $P_2$ is passively corrupted. The security holds evidently because there is no additional messages added to the view of $P_1$ and $P_2$ beyond those secured by key agreement, AES-128 hash and $\Pi_{mosn}$.

b) The $server$ is passively corrupted. The simulator $\textsf{Sim}$ learns $c=|X\cap Y|$ from $\mathcal{F}_{psa}$ and defines two vectors of random values of $\tilde{X}'=(\tilde{X}'_i)_{i\in[c]}$ and $\tilde{Y}'=(\tilde{Y}'_i)_{i\in[c]}$ such that there are $c$ identical elements. $\textsf{Sim}$ sends $\tilde{X}'$ and $\tilde{Y}'$ to server. The server cannot tell them from the genuine hash values which is ensured by security of AES-128 hash as discussed in Section~\ref{sec:lvl2}. Then, the security OF the following procedures relies on $\Pi_{mosn}$ which is proved to be secure in Theorem~\ref{thm:osnproof}. 
\end{proof}

\section{Conclusion}
PSA computes the inner join of distributed data in secret share format such that general analytics on the shared customers/users can be done using MPC. PSA can be used in various scenarios of privacy-enhancing data analytics, such as advertising and promotion campaign in e-commerce, collaborative money laundering detection across banks, disease prevention and control in healthcare, etc. In these cases, datasets from different entities may share a common group of customers and they expect to create value from the related data. 

In this work, we developed two PSA protocols to achieve two levels of privacy protection, i.e. ``single blinded'' and ``double blinded'' of common customers of the two datasets. We implemented our protocol and compared it with a HE-based solution to achieve the same level of privacy protection under the same security assumptions and the same experiment setup. According to our benchmarks, the proposed protocol is approximately $\times 100$ faster than its HE counterpart on a medium fast network (500 Mbps bandwidth), and the advantage is even bigger on a fast network. The speedup benefits from the computationally efficient operations in PSA. PSA also has its limits.

It is notable that many giant companies have a scale of customers in billions. As discussed in Section~\ref{sec:offline}, to handle datasets containing billions of records, the proposed protocol is capable to reduce the online run time to minutes by shifting majority of the work offline, while maintaining privacy and security.




\section*{Acknowledgments}
This research is supported by the National Research Foundation, Singapore and Infocomm Media Development Authority under its Trust Tech Funding Initiative and Strategic Capability Research Centres Funding Initiative. Any opinions, findings and conclusions or recommendations expressed in this material are those of the author(s) and do not reflect the views of National Research Foundation, Singapore and Infocomm Media Development Authority. We would also like to thank Ant Group CO., Ltd. for their valuable collaboration and support throughout this research.


\bibliography{ref}
\bibliographystyle{unsrt}


\begin{IEEEbiography}[{\includegraphics[width=1in,height=1.25in,clip,keepaspectratio]{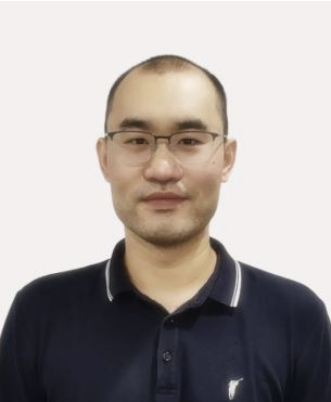}}]{Jiabo Wang}
Jiabo Wang received his B.Sc. degree from Harbin Engineering University, Harbin, China, in 2013, and M.Sc. degree from Tsinghua University, Beijing, China, in 2016, and Ph.D. degree from Imperial College London, London, UK, in 2021.
He is currently a Research Assistant Professor at SCRiPTS, Nanyang Technological University, Singapore. Prior to joining NTU, he was a Research Fellow of Tsinghua University, China (2021-2022).
His research interests include cryptography, privacy-preserving technologies. 
\end{IEEEbiography}

\begin{IEEEbiography} 
[{\includegraphics[width=1in,height=1.25in,clip,keepaspectratio]{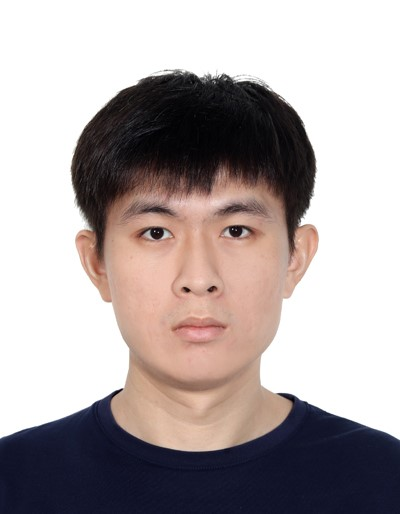}}]{Elmo Xuyun Huang} Elmo Xuyun Huang received his B.Sc and M.Sc from Nanyang Technological University, Singapore. He is now a PhD student in the College of Computing and Data Science at Nanyang Technological University, Singapore. His primary research interests lie in decentralized and distributed computing, as well as scalable privacy-preserving technologies. 
\end{IEEEbiography}

\begin{IEEEbiography}
[{\includegraphics[width=1in,height=1.25in,clip,keepaspectratio]{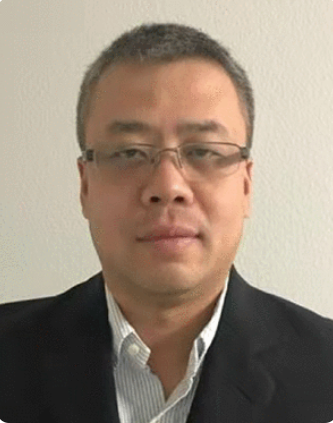}}]{Pu Duan} Pu Duan received the Ph.D. degree from Texas A$\&$M University, in 2011. He has been a 20 veteran on cryptography, information security, and networking security. He joined Cisco after obtaining his Ph.D. degree. At Cisco, he led the research and development of new cryptographic algorithms for TLS.13 on Cisco firewall product. He is currently working with the Secure Collaborative Laboratory (SCI), Ant Group, as a Senior Staff Engineer, leading the team on research and implementation of privacy-preserving technologies. He has published more than 20 papers on areas of cryptography, networking security, and system security.
\end{IEEEbiography}

\begin{IEEEbiography}[{\includegraphics[width=1in,height=1.25in,clip,keepaspectratio]{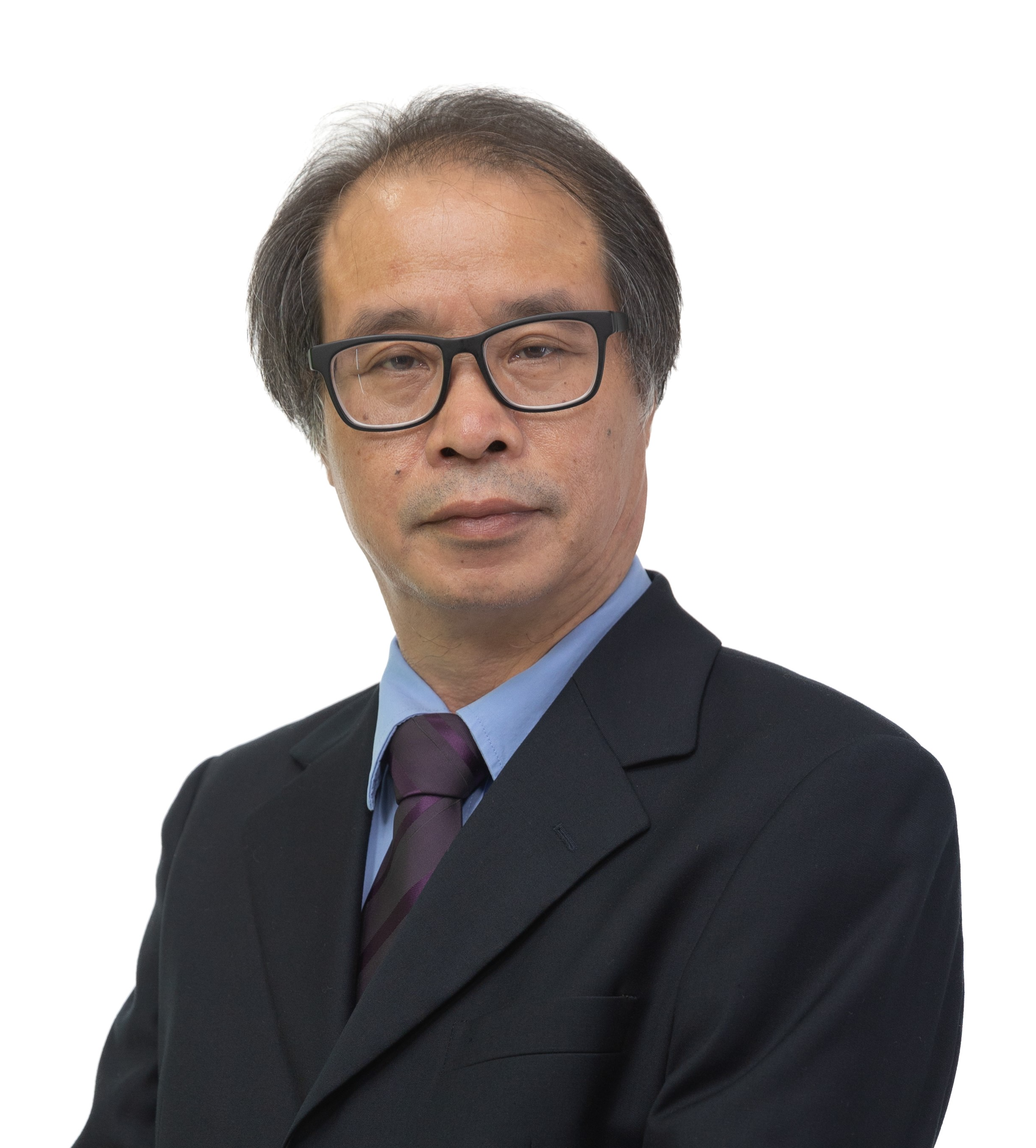}}]{Huaxiong Wang}
Huaxiong Wang received a Ph.D. in Mathematics from University of Haifa, Israel in 1996 and a Ph.D. in Computer Science from University of Wollongong, Australia in 2001. 
He has been with Nanyang Technological University in Singapore since 2006, where he is a Professor in the Division of Mathematical Sciences.  
Currently he is also the Co-Director of National Centre for Research in Digital Trust and the Deputy Director of Strategic Centre for Research in Privacy-Preserving Technologies and Systems at NTU.  
Prior to NTU, he held faculty positions at Macquarie University and University of Wollongong in Australia, and visiting positions at ENS de Lyon in France, City University of Hong Kong, National University of Singapore and Kobe University in Japan. 
His research interest is in cryptography and cybersecurity. He was the program co-chair of Asiacrypt 2020 and 2021.
\end{IEEEbiography}

\begin{IEEEbiography}[{\includegraphics[width=1in,height=1.25in,clip,keepaspectratio]{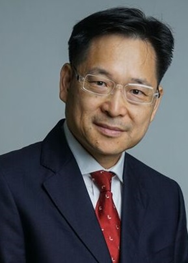}}]{Kwok-Yan Lam}
Kwok-Yan Lam (Senior Member, IEEE) received his B.Sc. degree (1st Class Hons.) from University of London, in 1987, and Ph.D. degree from University of Cambridge, in 1990. 
He is the Associate Vice President (Strategy and Partnerships) in the President’s Office, and Professor in the College of Computing and Data Science, at the Nanyang Technological University, Singapore. 
He is currently also the Executive Director of the National Centre for Research in Digital Trust, and Director of the Strategic Centre for Research in Privacy-Preserving Technologies and Systems (SCRiPTS). 
From August 2020, he is on part-time secondment to the INTERPOL as a Consultant at Cyber and New Technology Innovation. 
Prior to joining NTU, he has been a Professor of the Tsinghua University, PR China (2002–2010) and a faculty member of the National University of Singapore and the University of London since 1990. 
He was a Visiting Scientist at the Isaac Newton Institute, Cambridge University, and a Visiting Professor at the European Institute for Systems Security. 
In 1998, he received the Singapore Foundation Award from the Japanese Chamber of Commerce and Industry in recognition of his research and development achievement in information security in Singapore. 
He is the recipient of the Singapore Cybersecurity Hall of Fame Award in 2022. 
His research interests include Distributed Systems, Intelligent Systems, IoT Security, Distributed Protocols for Blockchain, Homeland Security and Cybersecurity.
\end{IEEEbiography}

\vfill
\clearpage
\pagenumbering{arabic}  
\appendices
\section{Construction of Bene\v{s} Network}\label{app:struBens}
In this section, we provide an illustration of how to construct generalized Bene\v{s} network to achieve a permutation for arbitrary $N$ using the method in \cite{CM97}.

\begin{figure}
    \centering
    \includegraphics[scale=0.75]{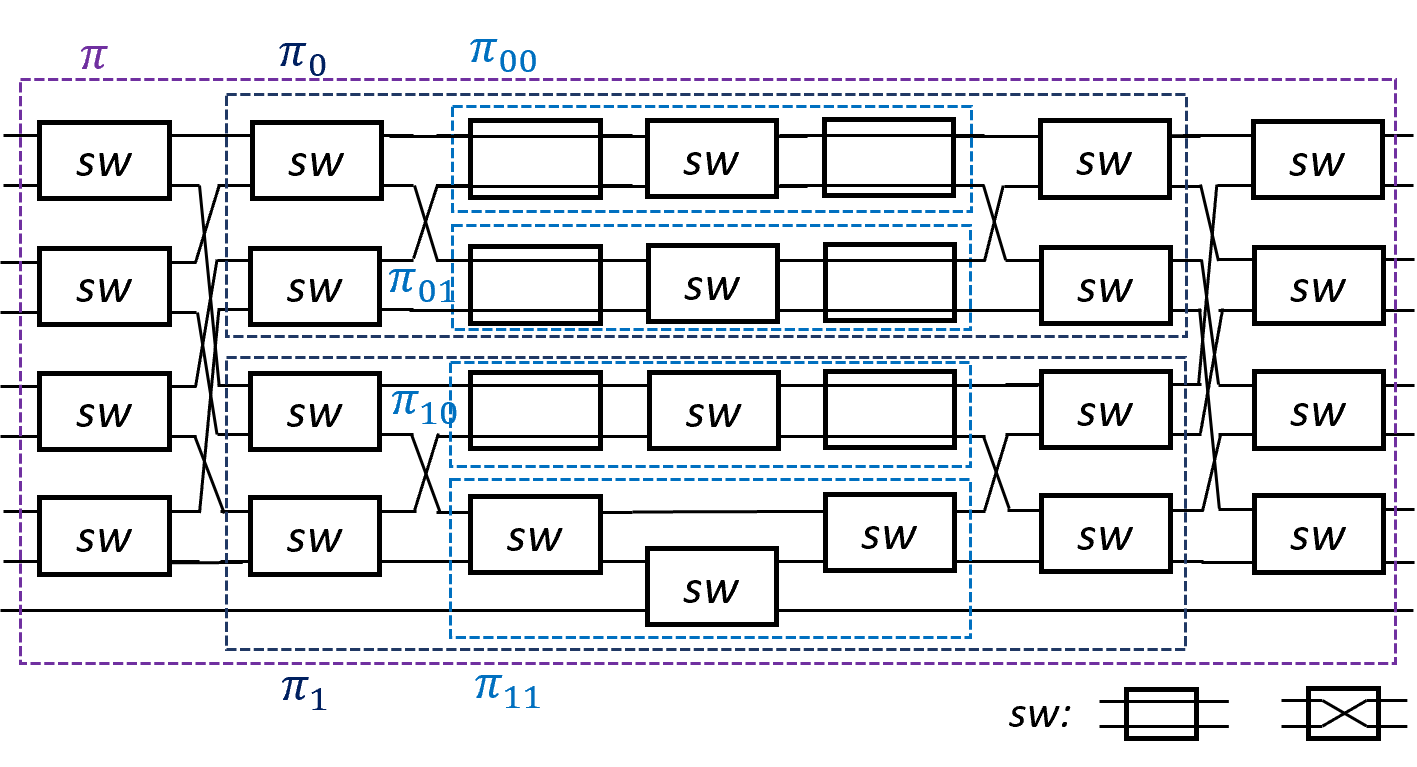}
    \caption{Generalized Bene\v{s} network for arbitrary $N$: an example for $N=9$, $\pi$ is recursively divided into sub-permutations $\pi_b,\pi_{bb}$ for binary $b$.}
    \label{fig:genrlBenes}
\end{figure}

\begin{lemma}\label{lem:generalBN}
Given any positive integer number $n\ge 2$, if we recursively split it into two additive integers $\lfloor n/2 \rfloor$ and $\lceil n/2 \rceil$, we will eventually reach a pair of addends being either 2 or 3. 
\end{lemma}
\begin{proof}
    The conclusion is evident, so proof is omitted.
\end{proof}
Given Lemma \ref{lem:generalBN}, one can generalize the Bene\v{s} network for arbitrary $N$ in a way that we recursively split a permutation of size $N$ into two sub-permutations of size $\lfloor N/2 \rfloor$ and $\lceil N/2 \rceil$ until a 2-by-2 or 3-by-3 sub-permutation is reached. An example for $N=9$ is depicted in Figure.~\ref{fig:genrlBenes}. We can observe that $\lfloor N/2 \rfloor$ many outgoing wires from the input layer are connected to the upper sub-network while the rest wires are connected to the lower sub-network. The recursion will terminate until a $2$-by-$2$ or $3$-by-$3$ sub-network in the middle is reached.
\begin{algorithm}
\caption{{Looping}($n$, $ptr1$, $ptr2$, $permIdx$, $src$, $dest$) for Bene\v{s} Network}
\begin{algorithmic}[1]
\item \textbf{Setup:}
A permutation $\pi: [N]\rightarrow[N]$ for arbitrary $N$; $n=\lceil \log_2N \rceil$; a global array $switches[:][:]$ of size $(2n-1)\times \lfloor N/2\rfloor$; a global vector $path[:]$ initialized as all $-1$s;
\item \textbf{Input:}    
  indices of current input and output layer $ptr1$ and $ptr2$, index $permIdx$ of current row of the array $switches$; $src[:]$ and $dest[:]$ are length-$N$ vectors before and after permutation.
\item \textbf{Output:} update the elements in the array $switches[:][:]$;
\Comment{\texttt{condition to return}}
\If{size of vector src == 2}
    \State $switches[ptr1][permIdx] = (src[0] != dest[0]$);
    \State return;
\EndIf
\State $N_1=\lfloor N/2 \rfloor$, $N_2=\lceil N/2 \rceil$; 
\Comment{\texttt{input/output for 1st sub-permutation}};
\State Declare two vectors $subsrc1[N_1]$ and $subdest1[N_1]$ 
\Comment{\texttt{input/output for 2nd sub-permutation}};
\State Declare two vectors $subsrc2[N_2]$ and $subdest2[N_2]$ 
\Comment{\texttt{permutation and inverse permutation}}
\State Initialize a vector $perm[:]$ of size $N$ to describe the permutation $\pi$ s.t. $dest[i]=src[perm[i]]$;
\State Initialize $invPerm[N]$ to describe $\pi^{-1}$ s.t. $src[i]=des[invPerm[i]]$;
\Comment{\texttt{Deep First Search to update} $path[:]$}
\State Let path[:] = -1s
\For{i =0 to N-1}
\If{path[i] == -1}
\State DFS(i);
\EndIf
\EndFor
\For{i = 0 to $\lfloor \frac{N-1}{2}\rfloor$}
\Comment{\texttt{Determine input layer gates} }
\State $switches[ptr1][permIdx + i]=path[2*i]]$
\Comment{\texttt{Determine} $subsrc1[:],subsrc2[:]$}
\If{$path[2*i]==0$ }
\State $subsrc1[i]=src[2*i]$;$subscr2[i]=src[2*i+1]$;
\Else
\State $subsrc1[i]=src[2*i+1]$;$subscr2[i]=src[2*i]$;
\EndIf
\EndFor

\For{i = 0 to $\lfloor\frac{N-1}{2} \rfloor$}
\Comment{\texttt{Determine output gates} $switches[ptr2][:]$}
\State $switches[ptr2][permIdx + i]=path[perm[2*i]]$
\Comment{\texttt{Determine} $subdest1[:],subdest2[:]$}
\If{$path[perm[2*i]]==0$ }
\State $subdest1[i]=src[perm[2*i]]$;
\State $subdest2[i]=src[perm[2*i+1]]$;
\Else
\State $subdest1[i]=src[perm[2*i+1]]$;
\State $subdest2[i]=src[perm[2*i]]$;
\EndIf
\EndFor

\State Looping($n-1$, $ptr1+1$, $ptr2-1$, $permIdx$, $subsrc1[N_1]$, $subdest1[N_1]$);
\State Looping($n-1$,  $ptr1+1$, $ptr2-1$, $permIdx + N/4$, $subsrc2[N_2]$, $subdest2[N_2]$);

\end{algorithmic}
\label{Algo:looping}
\end{algorithm}
\section{Subroutines }\label{app:osnroutines}
 Now we are ready to show the quasi-linear \textit{looping} algorithm to define the Bene\v{s} network to achieve arbitrary permutation $\pi: N\rightarrow N$ \cite{OT70}. We give the \textit{looping} algorithm for $2$-power $N$ and its subroutine in Algorithm~\ref{Algo:looping} and \ref{Algo:DFS}. The \textit{loop} for arbitrary $N$ is realized if we have a handle on the condition to terminate the recursion. In addition, for an odd $N$, the vertex $N-1$ and $\pi(N-1)$ of the undirected graph derived from the permutation $\pi$ will always be painted "1" because (a) at input layer, the $N$th wire is always connected to the lower sub-network (b) at output layer, the $N$th wire is also connected to the lower sub-network \cite{CM97}. For the sake of space limit, we remove that part from the pseudo-code.

We briefly describe the procedure of Algorithm \ref{Algo:looping} as follows. We define an array $switches$ of the same size of the network where each element indicate the ``color'' of a base switch gate. For a permutation $\pi:[N]\rightarrow[N]$, we denote by $src$ the vector of $[N]$ and denote by $dest$ its permutation. At the beginning, the indices $ptr1,ptr2$ point to the leftmost and rightmost column of array $switches$. Line 4 - Line 7: termination condition of the recursion. Line 11 - Line 18: describe the undirected graph by permutation $\pi$ and its inverse $\pi^{-1}$ then paint the vertices by \textit{deep first search}. Line 19 - Line 26: determine the ``color'' for gates at input layer. For a ``straight-through'' gate, the first input will be directed to an upper sub-network for a sub-permutation and the second input will be dicreted to the lower sub-network. It works the other way around for a ``crossover'' gate. Line 27 - 36: determine the ``color'' for the gates at output layer, and define the output of the two sub-permutations accordingly. Line 37 - 38: repeat the procedure for the two sub-permutations.

 \begin{algorithm}
\caption{DFS(i) for Bene\v{s} Network}
\begin{algorithmic}[1]
\State \textbf{Setup:} access to vector $perm[N]$ and $invPerm[N]$ for $\pi$ and $\pi^{-1}$;
\State \textbf{Input:} current node ID $i$ out of $0:N-1$
\State \textbf{Output:} updated vector $path$
\State Initialize a stack of pairs with a single pair $(0,0)$;
\While{stack is not empty}
\State Pop the top pair element from stack $(first,second)$;
\State $path[first] = second$;
\Comment{\texttt{One adjacent node}};
\State $AdjNode1 = dual(first)$;
\If{$AdjNode1$ is undetermined }
\State $stack.push( (AdjNode1, second\oplus1))$
\Comment{$dual(\cdot)$:\texttt{dual node (Def~\ref{def:dual})}}
\Comment{$second\oplus1$: \texttt{opposite color of} $second$}
\EndIf

\Comment{\texttt{The other adjacent node}};
\State $AdjNode2 = perm[dual(invPerm[first])]$;
\If{$AdjNode2$ is undetermined }
\State $stack.push( (AdjNode2, second\oplus1))$
\EndIf
\EndWhile
\end{algorithmic}
\label{Algo:DFS}
\end{algorithm}

\section{Oblivious Evaluation of Bene\v{s} Network}\label{app:evalBenes}
\begin{figure}
    \centering
    \includegraphics[scale=0.8]{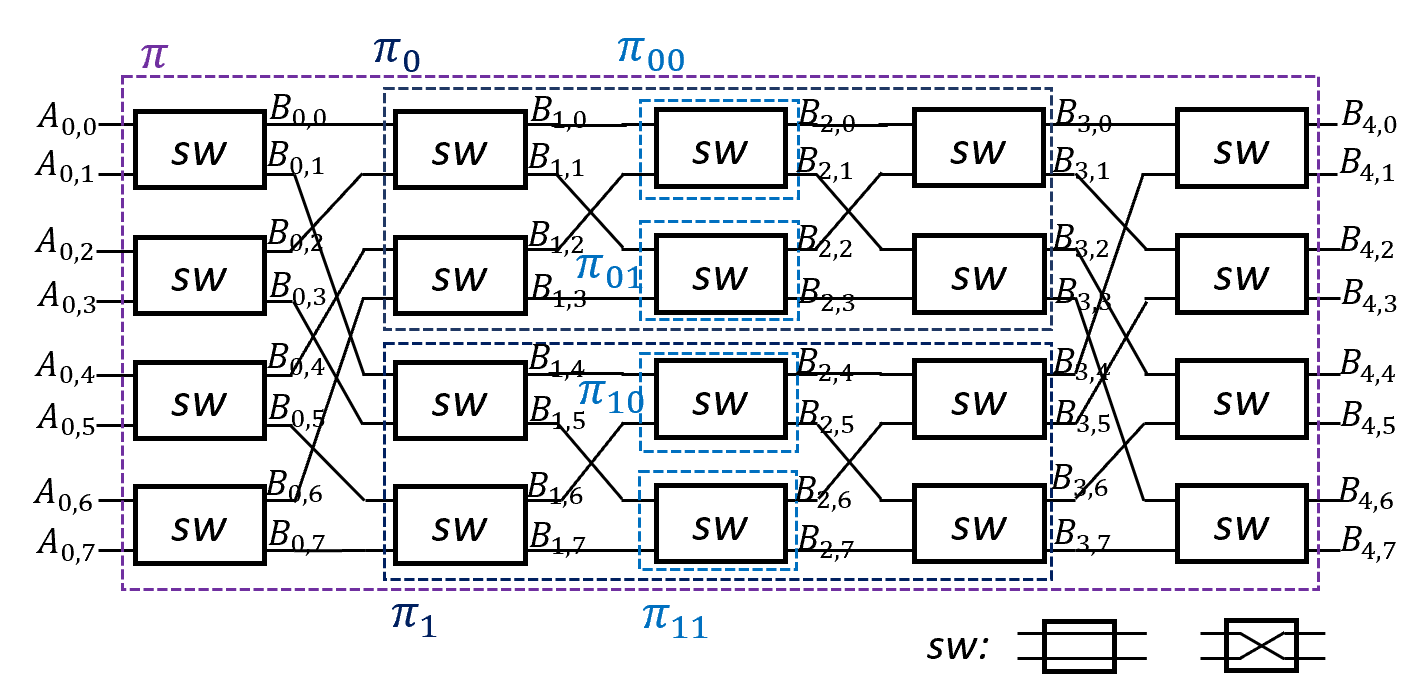}
    \caption{Oblivious evaluation of Bene\v{s} network: an example for $N=8$; $A_{i,j},B_{i,j}$ are random values assigned to I/O wires by OSN sender; $\pi$ is recursively divided into sub-permutations $\pi_b,\pi_{bb}$ for binary $b$ by OSN receiver.}
    \label{fig:senderOSNeval}
\end{figure}

\begin{algorithm}[!htb]
\caption{$evaluate(n,ptr1,permIdx,mskSrc)$: procedure of OSN receiver}
\begin{algorithmic}[1]
\item \textbf{Setup:}
A global array $switches$ of size $(2n-1)\times \lfloor N/2\rfloor$; OSN receiver saves output of 1-out-2 OT in an global 3-dimensional array $otVals$ of size $(2n-1)\times \lfloor N/2\rfloor \times 2$.
\item \textbf{Input:}    
  $n=\lceil \log N \rceil$, index of current column $ptr1$ of array $switches$, index $permIdx$ of current row of the array $switches$; a vector $mskSrc$;
\item \textbf{Output:} vector $mskScr$ 
\Comment{\texttt{condition to return}}
\If{$n$ == 1}
    \State $tmpBlocks = otVals[ptr1][permIdx]$;
    \State $mskSrc[0] = mskSrc[0]~\oplus~tmpBlocks[0]$;
    \State $mskSrc[1] = mskSrc[1]~\oplus~tmpBlocks[1]$;
    \If{$switches[ptr1][permIdx] == 1$}
    \State swap the values of $mskSrc[0],mskSrc[1]$;
    \EndIf
\EndIf
\State $N=2^n, N_1=\lfloor N/2 \rfloor$, $N_2=\lceil N/2 \rceil$; 
\Comment{\texttt{declare inputs for 2 sub-permutations}}
\State Declare $subMskSrc1[N_1]$ and $subMskSrc2[N_2]$;

\For{i = 0 to $\lfloor \frac{N-1}{2}\rfloor$}
\State $tmpBlocks = otVals[ptr1][permIdx+i]$
\State $mskSrc[2i] = mskSrc[i]~\oplus~tmpBlocks[0]$;
\State $mskSrc[2i+1] = mskSrc[i+1]~\oplus~tmpBlocks[1]$;
\Comment{\texttt{straight-through or crossover}}
\If{$switches[ptr2][permIdx + i] == 0$}
\State $subMskSrc1[i] = mskSrc[2i]$;
\State $subMskSrc2[i] = mskSrc[2i+1]$;
\Else
\State $subMskSrc1[i] = mskSrc[2i+1]$;
\State $subMskSrc2[i] = mskSrc[2i]$;
\EndIf
\EndFor
\Comment{\texttt{go for recursion}}
\State $eval(n-1,ptr1+1,permIdx,subMskSrc1)$;
\State $eval(n-1,ptr1+1,permIdx+N/4,subMskSrc2)$;
\Comment{\texttt{output at output layer}}
\For{i = 0 to $\lfloor \frac{N-1}{2}\rfloor$}
\State $s = switches[ptr1 + 2n-2][permIdx + i]$;
\If{$s == 0$}
\State $mskSrc[2i] = subMskSrc1[i]$;
\State $mskSrc[2i+1] = subMskSrc2[i]$;
\Else
\State $mskSrc[2i] = subMskSrc2[i]$;
\State $mskSrc[2i+1] = subMskSrc1[i]$;
\EndIf
\State $tmpBlocks = otVals[ptr1+2n-2][permIdx+i]$;
\State $mskSrc[2i] = mskSrc[2i]\oplus tmpBlocsk[s]$;
\State $mskSrc[2i+1] = mskSrc[2i+1]\oplus tmpBlocsk[1-s]$;
\Comment{\texttt{output saved in $mskSrc$}}
\EndFor
\end{algorithmic}
\label{Algo:receiverEval}
\end{algorithm}
There is an oblivious realization of shuffling using garbed circuit in \cite{huang2012private}. We employ the realization of OSN in \cite{GMR+21} which is a custom protocol for oblivious evaluation of a rearrangeable network. As shown in Figure~\ref{fig:funcOSN} there is a sender and a receiver in a OSN protocol. Because the ``hard-coded'' structures of the network is public known, the OSN sender assigns random values to every incoming and outgoing wire in the network as in Figure~\ref{fig:senderOSNeval}. The sender masks his input vector $U=(u_i)_{i\in[N]}$ with the random values $A_{0,:}$ assigned to the incoming wires at input layer. Then he sends the masked vector $(u_i \oplus A_{0,i})_{i\in[N]}$ to the receiver. After that, OSN sender and receiver invoke an OT extension \cite{iknp03} such that they produce $(2\lceil\log N \rceil -1)\times \lfloor N/2 \rfloor $ 1-out-2 OT instances, i.e. one instance per base switch gate. 
For a base switch gate at column $i$ and row $l$ in the network, assign $A_{i,j_0}$ and $A_{i,j_1}$ of the two random values associated with the outgoing wires of a gate it connects to from the left-hand side column, where $j_0,j_1\in J(l,2)$ are dual of each other with respect to characteristic $l$. Then, OSN sender synthesizes two messages as
$A_{i,j_0} \oplus B_{i,j_0}\| A_{i,j_1} \oplus B_{i,j_1} $ and $A_{i,j_0} \oplus B_{i,j_1}\| A_{i,j_1} \oplus B_{i,j_0}$ and inputs them to the 1-out-2 OT associated with this base switch gate. Eventually, sender outputs vector $B_{2n-2,:}$ for $n=\lceil\log N \rceil$. OSN receiver's procedure is given in Algorithm~\ref{Algo:receiverEval}. Benchmark is given in Table~\ref{tab:maskevaluate}.

\begin{table}[htbp]
    \centering
        \caption{Performance of Algorithm~\ref{Algo:receiverEval} \textit{evaluate}$(n,\cdot)$.}
    \begin{tabular}{|c|c|c|c|c|c|}
        \hline
    set size $n$ & $2^{14}$ & $2^{14}$ & $2^{16}$ & $2^{18}$ & $2^{20}$ \\
        \hline
        run time (ms) & $<1$ & $3$ & $11$ & $47$ & $211$ \\
        \hline
    \end{tabular}
    \label{tab:maskevaluate}
\end{table}
\section{Estimate the performance of the HE-based Protocol}\label{app:estimate}
We assume there are two dataset of $n$ records and a fraction $\alpha$ out of the $n$ records are those with common IDs. We estimate the performance of HE-based solution as follows. 
\begin{align*}
    &\text{comm} = 2n \times \text{key~size} \times (1.0+2\alpha),\\
    &\text{delay} = \text{comm} / bandwidth,  \\
    &\text{run~time} = 2(n \times ( (1+ \alpha) \times e  + \alpha \times s + \alpha \times d) + kg + \text{delay}),  \\
\end{align*}
    where $\text{key~size}=3072$ for 128-bit security, $e,s,d,kg$ are the time cost per encryption, subtraction, decryption, and key generation, respectively. There is a multiplier $2$ for comm and \text{run~time} because all benchmarking was done on a single thread and the two modified OSN instances are invoked sequentially.

Note that we get the estimation for HE-based protocol in Table~\ref{tab:TCinnerjoin} with the highest bandwidth giving it an advantage over our protocol. 


\end{document}